\documentclass[12pt]{article}
\usepackage{amsfonts}
\usepackage{psfig}
\usepackage{graphics}

\begin{document}

\newcommand{\R}{{\mathbb R}}
\newcommand{\C}{{\mathbb C}}
\newcommand{\Z}{{\mathbb Z}}
\newcommand{\be}{\begin{equation}}
\newcommand{\ee}{\end{equation}}
\newcommand{\bea}{\begin{eqnarray}}
\newcommand{\eea}{\end{eqnarray}}
\newcommand{\tb}{\tilde{\beta}}
\newcommand{\ta}{\tilde{a}}
\newcommand{\tal}{\tilde{\alpha}}
\newcommand{\ttau}{\tilde{\tau}}
\newcommand{\td}{\tilde{\delta}}
\newcommand{\cQ}{{\mathcal{Q}}}
\newcommand{\tQ}{{\tilde{\cQ}}}
\newcommand{\cF}{{\mathcal{F}}}
\newcommand{\cG}{{\mathcal{G}}}
\newcommand{\tG}{{\tilde{\cG}}}
\newcommand{\tN}{{\tilde{N}}}
\newcommand{\tF}{{\tilde{F}}}
\newcommand{\tV}{{\tilde{V}}} 
\newcommand{\cH}{{\mathcal{H}}}
\newcommand{\cX}{{\mathcal{X}}}  
\newcommand{\cY}{{\mathcal{Y}}}
\newcommand{\cR}{{\mathcal{R}}}
\newcommand{\cT}{{\mathcal{T}}}
\newcommand{\cA}{{\mathcal{A}}}  
\newcommand{\lt}{\frac{\Lambda}{3}}
\newcommand{\cM}{{\mathcal{M}}}

\newcommand{\nn}{\nonumber}
\newcommand{\tA}{\tilde{A}}
\newcommand{\tp}{\tilde{\phi}}
\newcommand{\e}{\varepsilon}
\newcommand{\mbin}{\mbox{in}}
\newcommand{\mbout}{\mbox{out}}

\newcommand{\az}{\alpha}
\newcommand{\bz}{\beta}
\newcommand{\cz}{\gamma}
\newcommand{\gz}{\gamma}
\newcommand{\dz}{\delta}
\newcommand{\ez}{\epsilon}
\newcommand{\kz}{\kappa}
\newcommand{\lz}{\lambda}
\newcommand{\na}{\nabla}

\newcommand{\pa}{p_\alpha}
\newcommand{\pb}{p_\beta}

\baselineskip0.6cm

\pagestyle{plain}
\pagenumbering{arabic}

\title{Cosmological Pair Production of Charged and Rotating Black Holes }

\author{I. S. Booth\footnote{ivan@avatar.uwaterloo.ca} 
    and R. B. Mann\footnote{mann@avatar.uwaterloo.ca}\\
        Department of Physics\\ 
        University of Waterloo\\   
        Waterloo, Ontario\\
        N2L 3G1}

\date{Jun 11, 1998\\}
\maketitle

\begin{abstract}
We investigate the general process of black hole pair creation in a
cosmological background, considering the creation of charged and rotating
black holes. We motivate the use of Kerr-Newmann-deSitter solutions to
investigate this process, showing how they arise from more general
C-metric type solutions that describe a pair of general black holes
accelerating away from each other in a cosmological background.
All possible KNdS-type spacetimes are classified and we examine
whether they may be considered to be in full thermodynamic
equilibrium. Instantons that mediate the creation of these space-times are
constructed and we see that they are necessarily complex due to regularity
requirements. Thus we argue that instantons need not always be real
Euclidean solutions to the Einstein equations. Finally, we calculate the
actions of these instantons and find that the standard action functional
must be modified to correctly take into account the effects of the
rotation. The resultant probabilities for the creation of the space-times
are found to be real and consistent with the interpretation that the
entropy of a charged and rotating black hole is the logarithm of the
number of its quantum states.
\end{abstract}

\section{Introduction}  
  
A considerable interest in black  
hole pair production has developed in recent years. Inspired by the well understood
particle pair production of quantum field theory (for example $2 \gamma  
\rightarrow e^+ + e^-$), theorists have investigated the  
possibility that a space-time with a source of excess energy will quantum  
tunnel into a space-time containing a pair of black holes. The earliest  
investigations considered pair creation due to background electromagnetic  
fields \cite{empair,hhr} but since then have been extended to include
pair creation due to  
cosmological vacuum energy
\cite{mann,othercosmo} ,cosmic strings \cite{strpair}, and 
domain walls \cite{dompair}.  These studies have repeatedly provided us
with evidence that the exponential of the entropy of a black hole does 
indeed correspond to the number of its quantum states. However to date 
they have all concentrated on the creation of non-rotating
black holes.

We seek here to extend black hole pair production
to include the creation of rotating black holes. As in the extant work we 
operate within the framework of the path integral formulation of 
quantum   
gravity, in which the probability amplitude for the creation of a pair of 
black  holes is approximated by $e^{-I_i}$, where  
$I_i$ is the action of a relevant instanton {\it i.e.} an imaginary-time  
solution to the field equations which interpolates between the states  
before and after the pair of black holes is produced.  
  
To study the pair creation of rotating black holes we proceed in   
stages. At the first stage, we obtain a solution to the Einstein-Maxwell   
equations that describes an appropriate 
pair of black holes accelerating 
away  from each other. Second, we construct an instanton that   
interpolates between this solution and some appropriate set of initial   
conditions. Finally, we calculate the action of this   
instanton, yielding the creation rates for the black hole pairs.   
  
The generalized $C$ metric of \cite{PlebDem} describes a pair of black 
holes  with opposite charge and rotation that are uniformly accelerating 
away from each other in a background with a cosmological constant. As 
such, it is the sought after 
solution to the Einstein-Maxwell  
equations describing the space-times that we wish to create. However, 
unless the rate of acceleration of the holes is carefully chosen to match that 
accounted for by the cosmological constant, these solutions will necessarily 
include cosmic strings or ``rods'' that provide the pressures/tensions 
necessary to account for the rest of the acceleration of the holes. Such 
structures manifest themselves as conical singularities in the metrics.
In this paper we will consider pair creation that is driven exclusively by
the 
cosmological vacuum energy and so start out by removing such 
singularities from the metrics.

Once a suitable class of solutions to the Einstein-Maxwell equations are in   
hand, we construct instantons to mediate the creation of said   
space-times. To describe the creation of rotating black holes we find it 
necessary to consider complex metrics so that the corresponding  
instantons may be successfully matched to physical solutions. This 
represents a departure from the 
usual viewpoint that the instanton actions in the path integral formulation 
must always be obtained from real, positive-definite (i.e. Euclidean)  
metrics,
obtained by  analytically continuing the parameters of the Lorentzian 
solution to imaginary values. For rotating black holes, this procedure
involves supplementing the analytic continuation $t\to it$ with the 
transformation $J\to iJ$, where $J$ is the real angular momentum 
\cite{GibH}. 
However (as noted previously \cite{BMY}) such an object has little to
do with a physical (Lorentzian) black hole, and we
find that this prescription cannot consistently define an instanton 
that will  mediate the pair creation of charged rotating black holes.
Despite our usage of complex   
metrics,  the Euclidean action associated with such instantons is 
always real, and the  probabilities are well defined.  In particular 
we find for all allowed 
instantons that the rate of pair creation is inversely proportional to 
the exponential of the total entropy  
of the black hole solutions, consistent with the expectation that black  
hole entropy is indeed associated with some (as yet unknown) degrees of   
freedom.

The outline of our paper is as follows. In section two 
we present a brief review of the path-integral formalism we employ. In 
section three we start by considering the general class of cosmological 
charged   rotating C-metrics, and show how   
a consideration of the removal of their conical singularities leads us to the  
Kerr-Newman-deSitter (KNdS) metrics. As a (useful) preliminary to the 
pair creation calculations, we then exhaustively classify the possible 
KNdS   
space-times. Since pair creation studies usually assume that the 
created space-times are in a state of thermodynamic equilibrium, we 
finish section three by  considering the KNdS space-times in that light.
  
In the fourth section we construct instantons that will mediate the  
creation of these space-times. According to the  
standard prescription for instanton construction we begin our  
constructions by analytically continuing the time to imaginary values. For  
rotating space-times this does not result in a real Euclidean  
metric, but rather in a complex metric, as noted above. We find that  
such complex metrics are demanded by the standard Euclidean  
formalism, and from them construct instantons for each  
of the space-times considered in section three.  
  
Finally, in section five we calculate actions for these  
instantons in order to estimate pair creation rates for the space-times of  
section three. The inclusion of rotation in these  
calculations is somewhat subtle, and we make use of the   
quasi-local formalism of Brown and York \cite{BY} in order to determine 
the 
functional form of the action that should be used to carry out these 
calculations. We obtain results that are consistent with   
earlier work on non-rotating black holes: that is, pair creation rates of
black   
holes are always suppressed relative to the  
creation rate of pure de Sitter space-time and such rates are 
inversely proportional  
to the exponential of one quarter of the sum of the areas of their black 
hole/cosmological   
horizons.  Since the creation rate is inversely proportional to   
the number of microstates, and since the area is proportional to the  
entropy of a given black hole space-time, this provides evidence  
that black hole entropy is indeed given by the logarithm of the number of  
microstates in the rotating case as well.

\section{The Path Integral Formalism} 
\label{formalism}  

In this section we briefly review the path integral formalism of quantum 
mechanics as it applies to a relativistic system with both gravitational 
and electromagnetic fields.  
 
Given a system whose classical evolution is governed by a lagrangian 
function $L$, a standard problem of quantum mechanics is to calculate the 
probability that the system passes from an initial 
state $X_1$ to a final state $X_2$. In a non-relativistic system each of 
these states may be described by specifying the state of the entire system 
at the corresponding instants of time $t_1$ and $t_2$. Of course for a 
relativistic system the concept of an instant of time is not so easily 
defined. For a four dimensional vacuum space-time with 
gravitational and electromagnetic fields the equivalent concept is to  
specify a 
three-manifold $\Sigma$ with Riemannian metric $h_{ij}$, a symmetric 
tensor field $K_{ij}$ (the extrinsic curvature, that physically describes 
how the spatial slice is evolving at that ``instant of time''), and two vector 
fields $E^i$ and $B^j$ that describe the electric and magnetic fields on 
$\Sigma$.  
In addition, in vacuo these four fields must satisfy the following constraint 
equations, 
\bea 
\cH &\equiv& ^{(3)}R + K^2 - K^{ij}K_{ij} - 2 (E^2 + B^2) = 0 
\label{MEC1}\\ 
\cH_i &\equiv& D_j K_i^{\; j} - D_i K - 2 \varepsilon_{ijk}E^j B^k = 0 
\label{MEC2}\\      
\cF_{el} &\equiv& D_j E^j = 0 \label{MC1}\\ 
\cF_{mg} &\equiv& D_j B^j = 0 \label{MC2}, 
\eea 
where $^{(3)}R$ is the Ricci scalar for $(\Sigma, h)$, $K = h^{ij}K_{ij}$, 
$E^2 = h_{ij}E^iE^j$, $B^2 = h_{ij}B^iB^j$, $D_j$ is the covariant 
derivative on $\Sigma$ that is compatible with $h_{ij}$, and 
$\varepsilon_{ijk}$ is the three dimensional Levi-Cevita tensor. These  
equations are the Einstein-Maxwell equations projected onto a spatial  
hypersurface. They ensure that the spatial slice along with its  
fields may be embedded in a 
larger four dimensional solution of the Einstein-Maxwell equations (in 
fact, if $\Sigma$ is a Cauchy surface they uniquely determine that 
solution via the evolution equations).  
Geometrically, the extrinsic curvature describes the shape of that 
embedding. 
 
The path integral approach to quantum mechanics then gives us a 
prescription for calculating the probability that the system will pass 
from $X_1$ to $X_2$. First, we must consider \underline{all} possible 
interpolations (or ``paths'') between the states (not just those that would be 
allowed by 
the classical evolution of the system). This  
means that we must consider all four-manifolds $M$, metric fields 
$g_{\az \bz}$ and electromagnetic field tensors $F_{\az \bz}$ on 
those manifolds such that the surfaces $\Sigma_1$ and $\Sigma_2$ and 
their 
accompanying fields, may be embedded in $M$ and its accompanying 
fields \footnote{In this context, we say that a three manifold $\Sigma$ and its 
accompanying fields $\{h_{ij}, K_{ij}, 
E_i, B_i\}$ may be embedded in the space-time $(M, g_{\az \bz}, F_{\az 
\bz})$ if there 
exists an embedding (in the differential topology sense), $\Phi:\Sigma 
\rightarrow M$ such that $\Phi^*(h_{ij}) = \left. h_{\az \bz} 
\right|_{\Sigma}$, $\Phi^*(K_{ij}) = \left. K_{\az \bz} \right|_{\Sigma}$, 
$\Phi^*(E_i) = \left. E_a \right|_\Sigma= \left. F_{\az \bz} u{^\bz} 
\right|_\Sigma$, and $\Phi^*(B_i) 
= \left. B_\az \right|_\Sigma = \left. -\frac{1}{2} \varepsilon_{\az  
\bz}^{\ \ \ 
\cz \dz} F_{\cz \dz} u^\bz \right|_\Sigma$. In the preceding $\Phi^*$ 
represents the 
appropriate mapping derived from $\Phi$ for the quantity being mapped, 
and 
$h_{\az \bz}$ and $K_{\az \bz}$ are respectively the induced metric on 
and 
extrinsic curvature of the surface $\Phi(\Sigma)$.}. We reiterate  
that the space-times $(M,g_{\az \bz}, F_{\az \bz})$ are not, in general, 
solutions to 
the Einstein-Maxwell equations.  
 
Second, the action  
\be
\label{actIM} 
I[M_{\Sigma_2 - \Sigma_1},g_{ab},F_{ab}] = \int_{M_{\Sigma_2 - 
\Sigma_1}} 
d^4 x \sqrt{-g} L(g_{ab},F_{ab}) + \mbox{ (boundary terms) }, 
\ee 
for each path must be calculated,  
where the integration is over all of $M$ between the two embedded 
surfaces $\Sigma_2$ and $\Sigma_1$ and the boundary terms are 
calculated 
on the boundaries of $M$ that are consistent with the boundaries of 
$\Sigma_1$ and $\Sigma_2$. They will be discussed in more detail later in 
this section and then in much more detail in section \ref{actions}. 
 
Third, each of these actions is used to assign a probability amplitude to 
its associated path. These amplitudes are then summed over all 
of the possible paths to give a net probability amplitude that 
the system passes from $X_1$ to $X_2$. This summation is represented as 
a 
functional integral over all of the possible manifold topologies, metrics, 
and vector 
potentials $A_\alpha$ (generating the field strength $F_{\alpha \beta}$)  
interpolating between the two surfaces. That is, 
\be 
\label{psi}  
\Psi_{12} = \int d[M]d[g]d[A] e^{-i I[M, g, F]}  \qquad .  
\ee 
 
Thus in principle, the probability that a space-time initially in a state 
$(\Sigma, h_{ij}, K_{ij},E_i,B_i)_1$ passes to a state $(\Sigma, h_{ij}, 
K_{ij},E_i,B_i)_2$ is proportional to $|\Psi_{12}|^2$ (we have not 
normalized the wave function). Unfortunately, as intuitively appealing as 
this formulation is, the integral (\ref{psi}) cannot be directly 
calculated. In the first place, there is no known way to define a measure for 
the 
integral. Second, even if such a measure were known, it seems 
quite likely that calculation of the integral would be impractical 
considering the uncountably infinite number of paths from $X_1$ to 
$X_2$. 
 
Fortunately there is a well-motivated simplifying assumption available. 
In analogy with flat-space calculations, it is argued \cite{OrigPathInt} 
that to lowest order in $\hbar$, the probability amplitude may be  
approximated (up to a normalization factor) by 
\be  
\label{pairc} 
\Psi_{12} \approx e^{-I_c}  
\ee  
where $I_c$ is the real action of a (not 
necessarily real) Riemannian solution to the Einstein-Maxwell equations 
that smoothly interpolates between the given initial and final conditions.  
Essentially, we have assumed that such a 
solution is a saddle point  
of the path integral. This solution (if it exists) is referred to as an 
instanton. The probability that such a tunnelling occurs is then
proportional to $|\Psi_{12}|^2 \approx e^{-2I_c}$.  Note that this 
interpretation requires that the action $I_c$ be real and positive, and 
that the fields satisfy the conditions (\ref{MEC1}--\ref{MC2}) so that 
the instanton smoothly matches onto the Lorentzian solution. 
 
We keep in mind that such Riemannian solutions are typically constructed 
by  analytically continuing the 
time coordinate $t$ of a Lorentzian solution to $i \tau$, and thereby 
changing the signature of the metric. As we have already noted, for 
rotating black 
holes such a continuation is not sufficient to turn a Lorentzian solution 
into a real Riemannian one. This issue will be dealt with in some detail 
in section four.     
 
Note also that it is not necessary for the instanton to have both 
initial and final conditions. If we can find a smooth Riemannian solution 
whose only boundary matches the final conditions $X_2$, then we can 
interpret 
the resultant probability as that for the creation of the 3-space $\Sigma$ 
from nothing. In that case we have chosen the initial boundary condition to 
be 
the no boundary condition of cosmology \cite{nobound}. 
 
As it stands, there is a gap in the above programme. Specifically, we have 
ignored the fact that the action (\ref{actIM}) which generates the 
classical equations of motion is not unique. Boundary terms consistent 
with the symmetries of the theory may always be added to the action 
without 
affecting the equations of motion. In  
order to make the right choice of action we must reinterpret the path 
integral 
formalism in terms of partition functions and thermodynamics.  We 
will postpone this issue until the computation of the action in 
section five. 
 
We turn next to a 
consideration of charged and rotating cosmological C-metrics.

\section{Rotating Black Hole Pairs}  
\label{KNdSsect}  

In this section we examine solutions that describe two black holes 
accelerating away from each other in a 
universe with a positive cosmological constant. In the first  
part, we shall examine the general solution describing such a situation and 
see  
how conservation of energy demands that the black hole acceleration rate 
be  
matched to the acceleration of the universe as a whole. We shall see that 
this  
concern forces us to consider Kerr-Newmann-deSitter (KNdS) space-times 
as the end states of black hole pair creation processes. As  
such we will classify the full range of these space-times. Finally, we will 
consider  
how quantum effects cause these solutions to evolve in time. 
  
\subsection{The Generalized C-metric}  
The well-known C-metric solution to the Einstein equations (first  
interpreted in \cite{KW})  describes  
a pair of uncharged and non-rotating black holes that are uniformly  
accelerating away from each other. In \cite{PlebDem} this metric was  
generalized to allow the holes to be charged and rotating, as well as to  
allow the inclusion of a cosmological   
constant and NUT parameter.  
  
In general, space-times of this type contain conical singularities.  
Physically these arise if the rate of acceleration of the black holes does  
not match the energy source available to accelerate them. Thus, in the  
cosmological case, if the black holes are accelerating faster or more  
slowly than the rest of the universe, conical singularities will exist.   
Physically, they may be interpreted as cosmic strings or ``rods'' that are  
pulling the black holes apart (or pushing them back together) so as to  
make them accelerate faster (or slower) than the  
rate of expansion of the universe. If we do not wish to consider such 
structures,  
then conservation of energy requires that we match the  
accelerations of the holes to the source of background energy and thereby  
eliminate the conical singularities.  
   
Here we demonstrate that one class of conical singularity free  
solutions of the generalized C-metric are the Kerr-Newmann-deSitter  
solutions. In other words, the KNdS solution may be viewed as a pair of 
oppositely charged and rotating black holes accelerating away from each 
other at a rate 
that matches the cosmological constant driven rate of acceleration of the 
universe as a 
whole. The generalized C-metric takes the form
\begin{equation}  
\label{PlebDemMetric}  
ds^2  = \frac{1}{(p-q)^2} \left[ \frac{1+p^2q^2}{P} dp^2   
+ \frac{P}{1+p^2q^2} \left( d \sigma - q^2 d \tau \right)^2   
- \frac{1+p^2q^2}{Q} d q^2   
+ \frac{Q}{1+p^2q^2} \left(p^2 d \sigma + d \tau \right)^2  
\right],  
\end{equation} 
with accompanying electromagnetic field defined by the vector potential 
\be 
A = - \frac{e_0 q (d \tau + p^2 d\sigma)}{1 + p^2 q^2} + \frac{g_0 p (d 
\sigma -  
q^2 d\tau)}{1 + p^2 q^2}, 
\ee 
where $p,q,\tau$, and $\sigma$ are coordinate functions,
\begin{equation}  
\label{Pp1}  
P(p) = (-\frac{\Lambda}{6} - g_0^2 + \gamma) + 2 n p - \epsilon   
p^2   
+ 2 m p^3 + (- \frac{\Lambda}{6} - e_0^2 - \gamma) p^4,  
\end{equation}  
and $Q(q) = P(q) + \frac{\Lambda}{3} (1 + q^4)$. $\Lambda$ is the   
cosmological constant,  $\gamma$ and $\epsilon$ are constants connected   
in a non-trivial way with rotation and acceleration, $e_0$ and $g_0$   
are linear multiples of electric and magnetic charge, and $m$ and   
$n$ are the respectively mass and the NUT parameter (up to a linear 
factor).   
This solution can be analytically extended across the coordinate singularity 
at  $p=q$, so that on the other side of $p=q$ we have a mirror image of the 
initial solution. Thus, if we view it as describing a pair of black holes, the 
two holes will be on opposite sides of that  $p=q$ hypersurface. 

In adaptations of this metric to more specific physical situations, the   
coordinate functions are associated with the more common spherical-type 
space-time  
coordinates as $q \leftrightarrow \frac{1}{r}$, $p   
\leftrightarrow \pa + \alpha \cos \theta$ for some constants $\alpha$ and   
$\pa$, $\sigma \leftrightarrow \phi$ and $\tau \leftrightarrow t$. Now   
in general, a periodic identification of $\sigma$ will introduce conical   
singularities at the roots of $P$. To avoid such singularities restrictions   
must be placed on the constants defining $P$.   
Defining $\pa$, $\pb$, $\alpha$, and $\beta$ so that the roots of $P(p)$ are   
at $\pa+\alpha$, $\pa-\alpha$, $\pb+i\beta$, and $\pb-i\beta$, we may write   
$P$ as  
\be 
\label{Pp2} 
P(p) = - C ([p-\pa]^2 - \alpha^2)([p-\pb]^2 + \beta^2),  
\ee  
where $C = -\frac{\Lambda}{6} - e_0^2 - \gamma$. 
We begin to specialize by assuming that only $\pa - \alpha$ and $\pa +   
\alpha$ are real roots, $\pa - \alpha < \pa + \alpha$ and $\pb,\beta \in \R$.   
Then there are only two real roots. Restricting  $p$ to lie between   
these two roots, we may reparameterize it as $p = \pa + \alpha \cos   
\theta$, where as usual $\theta \in [0, \pi]$. Then if $\pb=\pa$ (that is,   
$P(p)$ has an axis of symmetry along the line $p=\pa$), potential   
conical singularities at $\pa-\alpha$ or $\pa+\alpha$ may be   
simultaneously eliminated if we identify $\sigma$ with period $T =   
\frac{4 \pi}{P'(\pa - \alpha)}$ where $P' = \frac{d P}{d p}$.  
  
Next, we make the following extended series of coordinate   
transformations/definitions:  
\bea  
q &=& \frac{1}{\sqrt{\frac{\Lambda}{3}} \beta r}, \\   
\pa &=& \sqrt{\frac{\Lambda}{3}} \beta \tilde{\pa}, \\  
\pb &=& \sqrt{\frac{\Lambda}{3}} \beta \tilde{\pb}, \\ 
\alpha &=& \sqrt{\frac{\Lambda}{3}} \beta \tal, \\  
\chi^2 &=& 1 + \frac{\Lambda}{3} \tilde{\alpha}^2, \\  
\sigma &=& \frac{\phi}{\sqrt{\frac{\Lambda}{3}} C \beta^3   
\tilde{\alpha} \chi^2}, \\  
\tau &=& \frac{ t - \tilde{\alpha} \phi}{\sqrt{\frac{\Lambda}{3}} C   
\beta \chi^2}, \\   
\cH &=&  1 + \frac{\Lambda}{3} \tilde{\alpha}^2 \cos ^2 \theta,\\  
\cG &=& r^2 + (\tilde{\pa} + \tal \cos\theta)^2, \mbox{ and} \\  
\cQ(r)  &=& - \frac{\Lambda}{3 C} r^4 Q(q).  
\eea  
Further, equating (\ref{Pp1}) and (\ref{Pp2}) we note the   
following three equalities relating the two forms of $P$:  
\bea  
m &=& 2C\pa \\  
n &=& C\pa(2\pa^2 - \alpha^2 + \beta^2) ,\mbox{ and} \\  
g_0^2 + e_0^2 &=& C(1 + [\pa^2-\alpha^2][\pa^2 + \beta^2]) -   
\frac{\Lambda}{3}.  
\eea  
Then, after a significant amount of algebra, these transformations and 
equations will 
modify the metric (\ref{PlebDemMetric}) to become
\begin{equation}  
ds^2 = \frac{\Lambda}{3C (1 - \frac{\Lambda}{3} \beta^2 r [\tilde{\pa} +   
\tilde{\alpha} \cos \theta ])^2 } \left\{  
\begin{array}{lll}  
\frac{\cG}{\cH} d \theta^2 &+& \frac{\cH \sin^2 \theta}{\cG \chi^4}   
\left( \tal d t + [ r^2 + \tal^2] d \phi \right)^2 \\  
+ \frac{\cG}{\cQ} dr^2 &-& \frac{\cQ}{\cG \chi^4} \left( d t + \left[   
(\frac{\tilde{\pa}^2}{\tal} + 2 \tilde{\pa} \cos \theta ) -   
\tilde{\alpha} \sin^2 \theta \right] d \phi \right) ^2   
\end{array}  
\right\}.
\ee  
Setting $e_0 = \sqrt{\frac{\Lambda}{3}} E_0 \beta^2$, $g_0 =  
\sqrt{\frac{\Lambda}{3}} G_0 \beta^2$, and $\tilde{\pa} = M \beta^2$,  
we may write $\cQ$ as,  
\bea  
\cQ(r) &=& - \frac{\Lambda}{3} \left( \frac{1 - (E_0^2 + G_0^2) (M^2   
\beta^4 - \tal^2) ( 1 + \frac{\Lambda}{3} M^2 \beta^4) \beta^8}{1 -   
(E_0^2 + G_0^2) \beta^4} \right) r^4 \nn \\ 
&&- 2 \frac{\Lambda}{3}M \left( 1  +   
\frac{\Lambda}{3} (2 M^2 \beta^4 - \tal^2) \right)  \beta^2 r^3   
 + (1 + \frac{\Lambda}{3} (6M^2 \beta^4 - \alpha^2) )r^2 \nn
\\ &&- 2Mr +   
\frac{E_0^2 + G_0^2 + (\tal^2 - M^2 \beta^4)(1 + \frac{\Lambda}{3}   
M^2 \beta^4)}{1 + (E_0^2 + G_0^2) \beta^4}.  
\eea  
 
The $r^3$ term of the above is identified with the NUT parameter. 
If we wish to set this equal to zero but keep the mass parameter 
$M$ non-zero, then we must set one of $\beta$ or $1 + 
\frac{\Lambda}{3} (M^2 \beta^4 - \tal^2)$ to zero. Here we choose 
to take the limit as $\beta \rightarrow 0$ (choosing $1 + 
\frac{\Lambda}{3}
(M^2 \beta^4 - \tal^2) = 0$ results in a metric that is similar to but not
quite the KNdS metric - most notably it retains the leading conformal 
factor).
Then, if we replace 
$\tal$ with the more traditional symbol $a$ the metric becomes the 
standard Kerr-Newmann-deSitter metric (and similarly the vector 
potential $A$ becomes a vector potential that generates the 
associated electromagnetic field) which will be discussed in some 
detail in the rest of this section.  Thus, the KNdS metric describes two 
black holes in deSitter space that 
are accelerating away from each other due to the cosmological 
expansion of the universe.  

Before continuing, we pause to comment that there are 
other ways to eliminate the conical singularities in 
(\ref{PlebDemMetric}).  Although most yield the KNDS metric, some
will give rise to other space-times. These will not be considered in the 
present paper.
  
\subsection{The Basic Kerr-Newmann-deSitter Solution}  

In Boyer-Lindquist type coordinates, the Kerr-Newmann-deSitter metric 
takes the form \cite{MM}  
\bea  
\label{KNdS}  
ds^2  =  -\frac{\cQ}{\cG \chi^4} \left(  dt - a \sin^2\theta d\phi  
\right)^2 + \frac{\cG}{\cQ} dr^2  
      + \frac{\cG}{\cH} d\theta^2 + \frac{\cH \sin^2 \theta}{\cG \chi^4}  
\left( a dt - \left[ r^2 + a^2 \right] d\phi \right)^2,  
\eea  
where
\bea  
&& \cG \equiv r^2 + a^2 \cos^2 \theta, \; \; \;  \cH = 1 +  
\frac{\Lambda}{3}  
a^2  
\cos^2 \theta, \; \; \; \chi^2 = 1 + \frac{\Lambda}{3} a^2, \mbox{ and} \\   
&&\cQ = -\frac{\Lambda}{3} r^4 + \left( 1 - \frac{\Lambda}{3} a^2  
\right)  
r^2 - 2Mr + \left( a^2 + E_0^2 + G_0^2 \right).  \nn
\label{Qpoly}  
\eea  
The individual solutions are defined by the values of the   
parameters $\Lambda$, $a$, $M$, $E_0$, and $G_0$ which are   
respectively  
the cosmological constant (since we are interested  
in deSitter type space-times, we will assume that it is positive), the   
rotation parameter, the mass, and   
the effective electric and magnetic charge of the solution.  
Along with the electromagnetic field  
\be  
\label{EMfield}   
F = -\frac{1}{\cG^2 \chi^2} \left\{ X dr \wedge (dt - a \sin^2 \theta  
d\phi) + Y \sin  
\theta d\theta   
\wedge ( a dt - (r^2+a^2)d\phi) \right\},  
\ee  
where $X = E_0 \Gamma + 2aG_0 r \cos \theta $, $Y = G_0 \Gamma -   
2 a E_0 r   
\cos \theta$, and   
$ \Gamma = r^2 - a^2 \cos^2 \theta$, this metric is a solution to the  
Einstein-Maxwell equations. For future reference we note that a vector  
potential generating this field is
\be  
\label{EMpot}  
A = \frac{E_0 r}{\cG \chi^2} \left(dt  
 - a \sin^2 \theta d\phi \right) + \frac{G_0  
\cos \theta }{\cG \chi^2} \left( a dt - \left(  
r^2 + a^2 \right) d\phi \right).  
\ee  
  
The roots of the polynomial $\cQ$ correspond to horizons of the   
metric.   
As a quartic with real coefficients, $\cQ$ may have zero, two, or  
four real roots. We are interested in black hole space-times, and so shall  
assume that there are four real roots, and that three of them are  
positive. In ascending order the horizons corresponding to the positive 
roots   
are  
the inner and outer black hole horizons, and the cosmological horizon.    
  
Let the roots of $\cQ$ in increasing order be $d-\delta$, $d+\delta$,  
$e-\varepsilon $, and $e+\varepsilon$, where $e$ and $d$ are reals and  
$\varepsilon$ and $\delta$ are non-negative reals. The absence of a   
cubic  
term in $\cQ$ forces $d=-e$. Two further restrictions: 
\be  
\label{rootorder}  
\begin{array}{ccccl}  
0 & \leq & \varepsilon & < & e,  \mbox{  and}\\  
e & < & \delta & \leq & 2e - \varepsilon \nn  
\end{array} 
\ee
ensure that the roots are ordered as we have proposed.  
Then we may write $\cQ$ without loss of generality as  
\be  
\label{Qpoly2}  
\cQ = -\frac{\Lambda}{3} \left( (r-e)^2 - \varepsilon^2 \right) \left(  
(r+e)^2 - \delta^2 \right).  
\ee  
  
If all of the roots of $\cQ$ are distinct (we shall deal with the degenerate 
cases in 
subsections \ref{coldsect}, \ref{narsect}, and \ref{ucsect}), then by the 
standard 
Kruskal techniques the metric may be analytically continued through the 
horizons to 
obtain the maximal extension of the space-time \cite{cosmo}. Though this 
maximal
extension is infinite in extent, a variety of other global 
structures are possible if we choose to make periodic identifications
within it. In particular, if we demand that there be no closed 
time-like curves in the space-time and also wish to have two  
black holes in $t = \mbox{constant}$ spatial cross-sections, then the global 
structure is uniquely determined and is shown in figure \ref{PenReg}  
(for a two dimensional $\phi = \mbox{constant}$, $\theta = \frac{\pi}{2}$ 
cross 
section). The   
$t=\mbox{constant}$ spatial hypersurfaces are closed and each   
span the two black  
hole regions, cutting through the intersections of both the $r=r_c$  
and $r=r_o$ lines.     
The matching conditions are such that, in the  
spatial hypersurfaces,  the two holes have opposite  
spins as well as opposite charges. Thus, the net charge and net spin of  
the system are both zero. We note that it is not possible to periodically 
identify the 
space-time such that the spatial sections contain only a single black hole.  
\begin{figure}[t]
\centerline{\psfig{figure=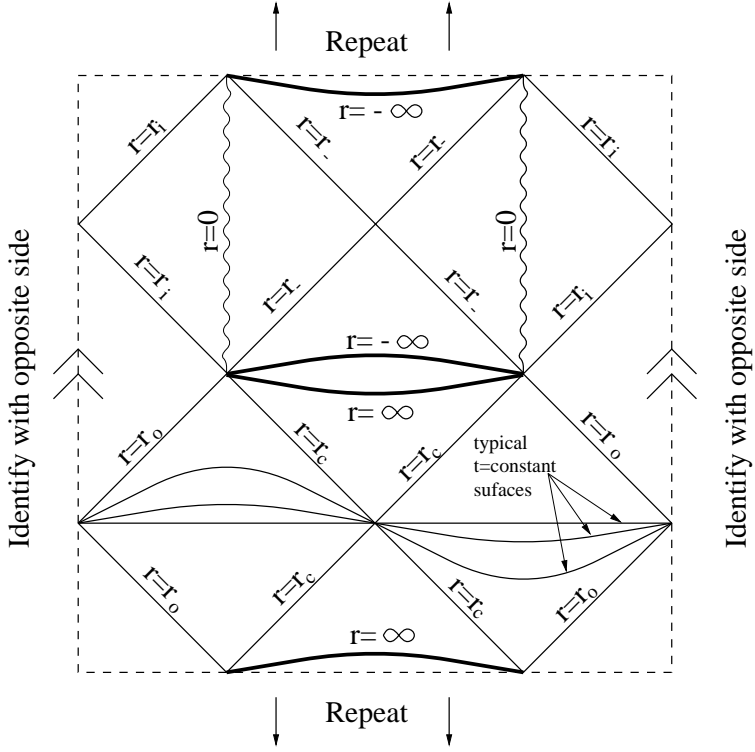,height=10cm,angle=270}}  
\caption{The global structure for the KNdS solutions - with periodic 
identifications to ensure that $t=\mbox{constant}$ hypersurfaces contain 
only two black holes. As indicated the figure is repeated vertically and 
periodically identified horizontally.
$r=r_c$ is the cosmological horizon, $r=r_o$ is the outer  
black hole horizon, and $r=r_i$ is the inner black hole horizon. The  wavy   
lines at $r=0$ represent the ring singularity found there for $a \neq 0$.  
If $a = 0$ then this singularity may not be avoided and the space-time  
cuts off at $r=0$. Otherwise the singularity may be bypassed and we may  
proceed to negative values of $r$. $r=r_-$ is the (negative) fourth root of 
$\cQ$.}   
\label{PenReg}  
\end{figure}  
  
Next, we consider the range of the parameters for which  
our  
conditions on the roots of $\cQ$ will hold.  
  
\subsection{The Allowed Range of the KNdS Solutions}  
The requirement that $\cQ$ has three positive real roots enforces   
restrictions on the allowed values of the physical parameters $a$, $M$,  
$E_0$, and $G_0$. $\cQ$ is a quartic, and so in principal we may   
solve it  
exactly and decide under what circumstances it has four real roots. In  
practice however, the exact solution to a quartic is too messy to work  
with. Thus, we tackle the problem in reverse. We will first determine   
the  
allowed ranges of the $\cQ$ structure parameters $e$, $\delta$, and  
$\varepsilon$, and then use these to parameterize the allowed range of   
the  
physically meaningful parameters $a$, $M$, $E_0$, and $G_0$.  
  
Matching (\ref{Qpoly}) with (\ref{Qpoly2}) we obtain expressions  
for the physical parameters in terms of the structure parameters:   
\bea   
\label{a2}   
a^2 & = & \frac{3}{\Lambda} - \delta^2 - \varepsilon^2 - 2e^2, \\   
\label{M}   
M & = & \frac{\Lambda}{3} (\delta^2 - \varepsilon^2) e, \mbox{ and}   
\\  
\label{Q2}   
E_0^2 + G_0^2 & = & \lt (\delta^2 - e^2)(e^2 - \e^2) + (\delta^2 +   
\e^2 +  
2e^2) - \frac{3}{\Lambda}.  
\eea   
  
Requiring that each of these parameters be non-negative will impose  
further restrictions (beyond the root ordering conditions  
(\ref{rootorder}))  
on the allowed ranges of $e$, $\varepsilon$, and $\delta$.   
Requiring that $a^2 \geq 0$ we obtain the condition   
\be   
\label{pc1} \frac{3}{\Lambda} - \delta^2 - \varepsilon^2 -  
2e^2 \geq 0.  
\ee   
$M$ will automatically be non-negative because of the root-ordering 
conditions (\ref{rootorder})  
while requiring that $E_0^2+G_0^2 \geq 0$ we obtain   
\be   
\label{pc2}  
\frac{\Lambda}{3} (\delta^2 - e^2)(e^2 - \varepsilon^2) + (\delta^2 +  
\varepsilon^2 + 2e^2) - \frac{3}{\Lambda} \geq 0.    
\ee  
In order to disentangle these structure parameters we rescale them   
as follows.
$\Lambda$ and $e$ are non-zero so we may define $\Delta$, $E$, and  
$X$ by:
\be  
\label{defs}  
\delta \equiv \Delta e, \mbox{ } \varepsilon \equiv E e, \mbox{  
and } e \equiv \sqrt{ \frac{3}{\Lambda} } X.  
\ee  
Then, the conditions (\ref{pc1}) and (\ref{pc2}) become respectively,  
\bea  
\label{c1}  
&& 1 - (\Delta^2 + E^2 + 2) X^2 \geq 0, \mbox{ and } \\  
\label{c2}   
&& (\Delta^2 - 1)(1 - E^2) X^4 + (\Delta^2 + E^2 + 2) X^2 - 1 \geq 0  
\eea  
The first of these provides an upper bound on the allowed range $X$  
for given values of $\Delta$ and $E$. $a^2 \geq 0$ if and only if  
\be  
X \leq X_U \equiv \frac{1}{\sqrt{2 + \Delta^2 + E^2}}.  
\ee 
  
In the meantime, (\ref{c2}) is quadratic in $X^2$ and so may be easily   
solved.  
It turns  
out that over the allowed ranges of $\Delta$ and $E$, it has only  
one positive real root. Further, it is upward  
opening, and therefore the positive real root provides a lower bound   
for  
the allowed values of $X$. $E_0^2 + G_0^2 \geq 0$ is and only if  
\be  
X \geq X_L \equiv \sqrt{  \frac{  -(\Delta^2 + E^2 + 2) +  
\sqrt{8(E^2+\Delta^2)+(E^2-\Delta^2)^2 }}{2 (\Delta^2 - 1)(1 - E^2)   
}}.  
\ee  
  
On plotting $X_U$ and $X_L$ we find that for $0 \leq E \leq 1$ and $1   
\leq  
\Delta \leq 2$, $X_L \leq X_U$ and so there exists a non-zero range   
for $X$   
for all the possible values of $E$ and $\Delta$. With this range of  
allowed values for $X$ in hand, we now have  
a parameterization for all the possible KNdS black hole solutions. The  
parameterization is given by the restrictions  
\be  
1 < \Delta \leq 2, \mbox{ } 0 \leq E < 2 - \Delta, \mbox{ and } X_L  
\leq X \leq X_U,  
\ee  
the definitions (\ref{defs}), and the expressions (\ref{a2})-(\ref{Q2}).  
  
These ranges are shown in figure \ref{allowedRange}. In that figure 
the allowed parameter range of KNdS space-times is the region bounded by 
the five 
sheets defined by $a = 0$, $M = 0$,  
$E_0^2+G_0^2 =0$, $E=0$, and  
$E = 2 - \Delta$. The last two conditions are respectively cold black hole 
space-time    
where the inner and outer black hole horizons coincide 
and a Nariai-type space time where the outer black hole horizon coincides   
with the cosmological horizon (we shall soon see that 
this apparent degeneracy  of the metric is an artifact 
of the coordinate system and that the distance between the two 
horizons remains finite and non-zero throughout  
the limiting process). The intersection of the Nariai and cold   
sheets is referred to as the ultracold solution. This nomenclature is taken 
from  
the corresponding non-rotating instantons discussed in \cite{mann}, and 
will be used 
throughout this work. A special case of solutions labelled the lukewarm 
solutions is 
also shown in the figures. It will be discussed in subsections \ref{equil} 
and \ref{luke}.  
\begin{figure}  
\centerline{\psfig{figure=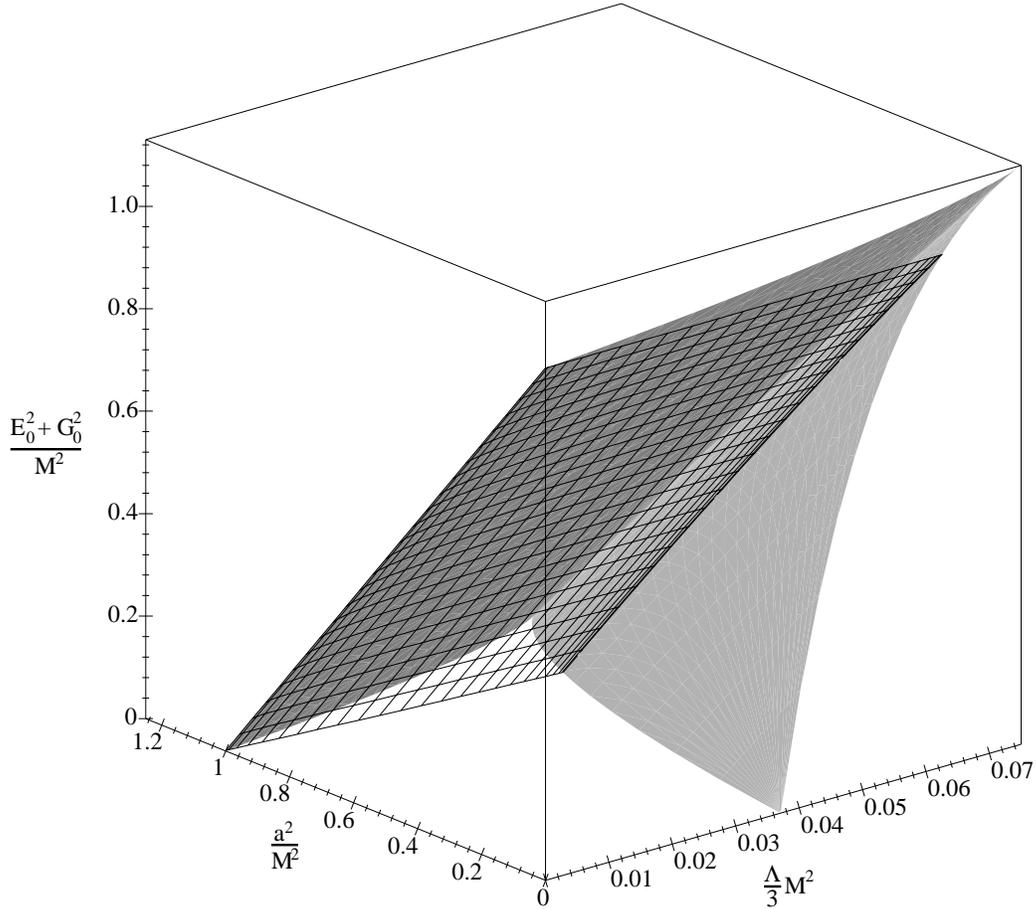,height=13cm}}  
\caption{The allowed range of the KNdS parameters. The range is bounded 
by the planes $M=0$, $a^2 = 0$, $E_0^2 + G_0^2 = 0$, the cold solutions 
(the darkest sheet) and the rotating Nariai solutions (the lighter gray sheet).
Also shown as a meshed sheet are the lukewarm solutions.}
\label{allowedRange}   
\end{figure}  
  
We have now established the range of KNdS solutions allowed by the  
structure of the polynomial $\cQ$. It remains to be demonstrated that   
the full 
range is realizable as a set of well defined metrics. In particular the  
current coordinate representation of the metric breaks down in the   
Nariai  
($\varepsilon \rightarrow 0$, $\delta \neq 0$)  and ultracold  
($\varepsilon  
\rightarrow 0$, $\delta \rightarrow 2e - \varepsilon$) cases. In the  
following  
three subsections we shall consider how these various limits may be  
achieved, and we will further consider how the limiting processes   
affect  
the global structure of the space-times. We begin with  
the cold limit ($\delta = 2e - \varepsilon$, $\varepsilon \neq 0$).   
  
\subsection{The Cold Limit}  
\label{coldsect}   
This limit may be taken without having to make  
any changes to the coordinate system. Therefore, the metric keeps the   
form  
(\ref{KNdS}) and the electromagnetic field and potential remain as  
(\ref{EMfield}) and (\ref{EMpot}) respectively. The physical   
parameters  
are given by:   
\bea   
a^2 &=& \frac{3}{\Lambda} - 2(3e^2 - 2 \varepsilon e +  
\varepsilon^2) \\   
M &=& \frac{4 \Lambda}{3} e^2 (e-\varepsilon), \mbox{  
and} \\   
E_0^2 + G_0^2 &=& \lt (3e - \e)(e - \e)^2 (e+\e) + 2(3e^2 - 2e\e +   
\e^2)  
- \frac{3}{\Lambda},   
\eea   
where the range  
of the parameters is limited by the relations   
\bea   
0 <& E &< 1, \mbox{ and}\\   
\sqrt{\frac{-3 + 2E - E^2 + 2 \sqrt{3 - 4E + 2E^2}}{(3-E)(1+E)(1-  E)^2}}  
\leq & X & \leq \frac{1}{\sqrt{2(E^2 - 2E +3)}}.   
\eea   
As before, $e =  
\sqrt{ \frac{3}{\Lambda} } X$, and $\varepsilon = E e$.   
  
In this space-time, the double horizon of the black hole recedes to an  
infinite proper distance from all other parts of the space-time.  Thus, the 
global  
structure of the space-time changes - in particular, the region inside the  
black hole is cut off from the rest of the space-time. Again we choose   
the global structure so that the space-time contains two (in this case   
extreme) black holes. This structure is shown in figure \ref{PenCold}. Note 
that  
in this case, the $t=\mbox{constant}$ hypersurfaces consist of two   
extreme black holes, and so are not closed as they were in the lukewarm 
case   
(the horizons recede to infinite proper distance from all other points in the  
space-time).
  
Finally, we note for the cases where $a=0$, this solution reduces to the  
cold solutions discussed in \cite{mann}.   
  
\begin{figure}  
\centerline{\psfig{figure=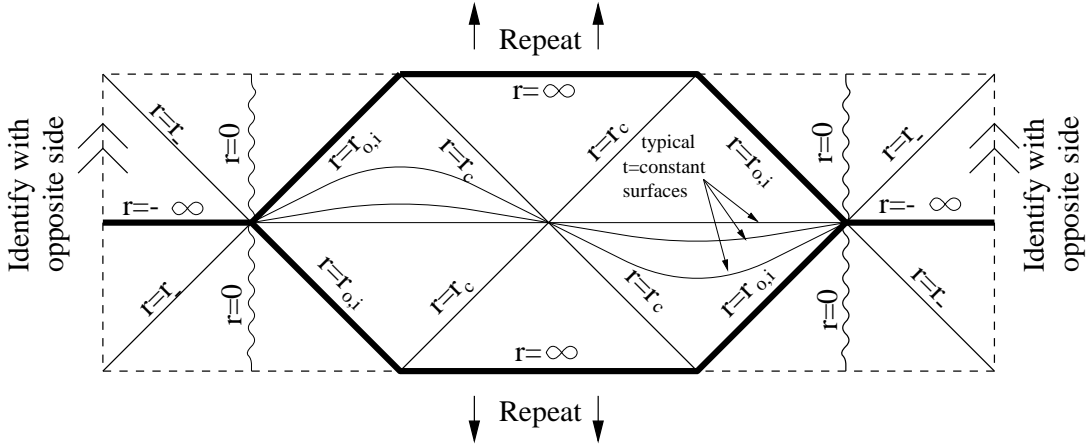,height=6cm,angle=270}}  
\caption{The Penrose diagram for a two hole cold KNdS space-time.  
Opposite sides of the rectangle are identified. $r=r_c$   
is the cosmological horizon and $r=r_{o,i}$ is the double black hole  
horizon. If $a=0$, then the space-time cuts off at the singularity at   
$r=0$. Otherwise, we may pass through the ring singularity to the  
negative values of $r$, including $r_-$, the fourth root of $\cQ$.}  
\label{PenCold}  
\end{figure}  
  
\subsection{The Nariai Limit}   
\label{narsect}   
The current coordinate  
system breaks down in the $\varepsilon = 0$ limit. Specifically, for  
$\varepsilon = 0$, $r=e$ (becomes a constant), and $\cQ=0$, so the  
coordinate system becomes degenerate, and the metric ill-defined.  
These problems may easily be avoided however, if we make the  
transformations:   
\bea   
r &=& e + \varepsilon \rho. \\   
\phi &=& \varphi + \frac{a}{e^2 + a^2} t, \mbox{ and}\\   
t &=& \frac{(e^2+a^2)\chi^2}{\varepsilon} \tau.    
\eea   
Then, the $\varepsilon \rightarrow 0$ limit may  
be taken without hindrance, and the metric becomes   
\be ds^2 = - \tQ \cG d \tau^2 + \frac{\cG}{\tQ} d \rho^2 +   
\frac{\cG}{\cH}  
d\theta^2  
     + \frac{\cH \sin^2\theta}{\cG} \left( 2 a e \rho  d \tau +  
\frac{e^2+a^2}{\chi^2} d\varphi\right)^2,  
\ee     
while the electromagnetic  
field becomes,  
\be  
F = \frac{-X}{\cG} d\rho \wedge d\tau + \frac{Y \sin \theta}{\cG^2}   
d\theta \wedge \left( 2ae\rho   
d\tau + \frac{e^2 + a^2}{\chi^2} d\varphi \right).
\ee  
A electromagnetic potential generating this is
\be  
A = - E_0 \frac{(e^2 - a^2)}{e^2 + a^2} \rho d\tau - \frac{a E_0 e  
\sin^2 \theta + G_0 (e^2+a^2) \cos \theta}{\cG (e^2+a^2) }  \left( 2  
a e \rho d\tau + \frac{e^2 + a^2}{\chi^2} d \varphi \right).   
\ee
In the above, $\tQ = \frac{\Lambda}{3} (2e-\delta)(1-  
\rho^2)(2e+\delta)$, $\cG   
= e^2 + a^2 \cos^2 \theta$, $\Gamma = e^2 - a^2 \cos^2 \theta$, $X =   
E_0 \Gamma + 2a G_0 e   
\cos \theta$, and $Y = G_0 \Gamma - 2 a E_0 e \cos \theta$.  
Note that the above potential is not the simply ($\ref{EMpot}$) under
the coordinate
transformation as the $A$ generated in that way diverges when
$\varepsilon \rightarrow 0$. The divergence is removed (and the above
result obtained) if we make the gauge transformation $A \rightarrow A
- \frac{E_0 e}{\varepsilon} d \tau$ before we take the coordinate
transformation and limit.
  
The physical parameters are given in terms of $e$ and $\delta$ as:  
\bea  
a^2 &=& \frac{3}{\Lambda} - 2e^2 - \delta^2, \\  
M &=& \frac{\Lambda}{3} \delta^2 e, \mbox{ and}\\  
E_0^2 + G_0^2 &=& \lt(\delta^2 - e^2)e^2 + (2e^2 + \delta^2) -  
\frac{3}{\Lambda},  
\eea  
and the allowed ranges of $e = \sqrt{\frac{3}{\Lambda}} X$ and  
$\delta = \Delta e$ are given by
\bea  
1 <& \Delta &\leq 2 \mbox{ and} \\  
\sqrt{\frac{-(\Delta^2+2)+\Delta \sqrt{\Delta^2 + 8}}{2(\Delta^2-1)}}   
\leq   
& X & \leq \frac{1}{\sqrt{ 2 + \Delta^2}}.  
\eea  
  
We note that the Nariai solution is no longer a black hole solution.  
Extending the metric through the horizons by the standard Kruskal  
techniques, we obtain the Penrose diagram of figure \ref{PenNar} for   
the  
($\tau$,$\rho$) sector. Note that there is no longer a singularity at  
finite distance beyond either of the horizons, and so this is no longer a  
black hole space-time. In fact, the diagram is the same as that for two  
dimensional deSitter space. If there were no rotation ($a=0$), then this  
space-time would just be the direct product of two dimensional deSitter  
space, and a two sphere of fixed radius. With rotation, of course the  
situation is not so simple.  
\begin{figure}  
\centerline{\psfig{figure=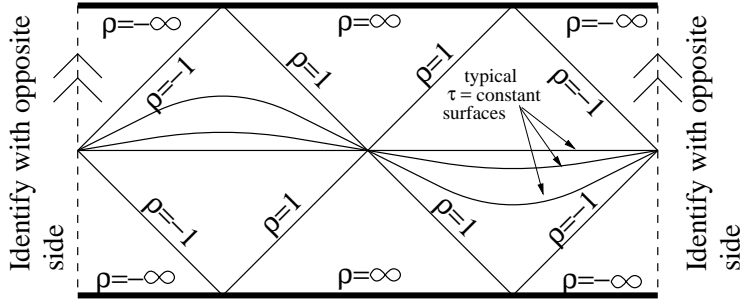,height=4cm,angle=270}}  
\caption{The Penrose diagram for the Nariai limit space-time. 
$\rho = \pm 1$ are the two cosmological horizons.}  
\label{PenNar}  
\end{figure}  
  If $a=0$, and we make the coordinate transformation $\rho = \cos   
\chi$, then this solution reduces to the non-rotating charged Nariai solution  
considered in \cite{mann}.  
  
Even though the Nariai solution is not a black hole solution itself,  
it was shown in \cite{ginsper} that an uncharged, non-rotating Nariai  
solution is unstable with respect to quantum tunnelling into an almost-
Nariai 
Schwarzschild-deSitter space-time. It is usually argued     
\cite{bousso} that this tunnelling carries over analogously with the  
inclusion of charge and rotation - in which case Nariai solutions  
decay into near Nariai KNdS space-times. Thus, in the future sections   
of  
this paper where we study black hole  
pair creation this solution will remain of interest. Then, a route to  
black  
hole pair creation will be to create a Nariai space-time and then let it  
decay into a black hole pair.  
  
\subsection{The Ultracold Limits}  
\label{ucsect}  
Finally we consider the ultracold limits where both $\varepsilon  
\rightarrow  
0$ and $\delta \rightarrow 2e-\varepsilon$. It turns out that there are  
two such limits which we shall label the ultracold I and II limits. In  
this subsection we shall only demonstrate how they may  
be reached from the Nariai limit. Similar coordinate  
transformations (which sometimes must be iterated two or three times)  
allow us to reach the same two limits both from the cold limit, and,  
taking $\delta \rightarrow 2e-\varepsilon$ and $\varepsilon \rightarrow   
0$  
simultaneously, straight from the non-extreme standard KNdS form of  
the metric. We deal with the two cases separately.  
 
\underline{Ultracold I:} 
Making the transformations, 
\bea 
\rho &=& \eta - \eta k (2e-\delta) R, \\ 
\varphi &=& \Phi - 2 \eta \frac{ae \chi^2 \tau}{e^2+a^2}, \mbox{ 
and}\\ 
\tau &=& \frac{ \eta T}{k (2e-\delta)}, 
\eea   
where $\eta = \pm 1$, and $k = 8 \lt e$, and taking the 
limit as $\delta \rightarrow 2e$ we obtain, 
\be 
ds^2 = -\cG R dT^2 + \frac{\cG}{R} dR^2 + \frac{\cG}{\cH} 
d\theta^2 + 
\frac{\cH}{\cG}\sin^2\theta \left( 2 a e R dT + 
\frac{e^2+a^2}{\chi^2} 
d \Phi\right)^2. 
\ee 
The electromagnetic field and potential become, 
\be    
\label{UCF} 
F = \frac{-X}{\cG} dR \wedge dT + \frac{Y \sin \theta}{\cG^2} 
d\theta \wedge \left( 2 a e R dT + \frac{e^2 + a^2}{\chi^2} d\Phi 
\right), 
\ee 
and, 
\be 
\label{UCA} 
A = -E_0 \frac{e^2-a^2}{e^2+a^2} R dT - \frac{a E_0 e \sin^2 \theta 
+    
G_0 (e^2+a^2) \cos \theta}{\cG (e^2+a^2)} \left( 2 a e R dT + 
\frac{e^2+a^2}{\chi^2} d \Phi \right). 
\ee 
$R \in (0, \infty)$, $T \in (-\infty, \infty)$, 
$\theta \in [0, \pi]$, and $\Phi$ inherits a $2\pi$ periodicity from its 
predecessors. 
$\cG$, $\cH$, $\chi^2$, $X$, and $Y$ all retain their old 
definitions. Note that the EM potential and field have retained their Nariai 
form. 
 
The ($R$,$T$) sector of the space-time is conformally the 
same   
as the Rindler space-time (which of course is actually a sector 
of two dimensional Minkowski space). The Rindler horizon is at 
$R=0$ and as this is the only horizon, the space does not contain black 
holes. 
Before giving the 
parameterization of this solution, 
we consider the transformations leading to the ultracold II case. 
 
\underline{Ultracold II:}  
Making the transformations,  
\bea  
\rho &=& b + k \sqrt{2e-\delta} R, \\  
\varphi &=& \Phi - 2 \frac{aeb \chi^2 \tau}{e^2+a^2}, \mbox{ and}\\  
\tau &=& \frac{T}{k \sqrt{2e-\delta}},  
\eea  
where $b \neq \pm 1$, and $k = 2\sqrt{ \lt (1-b^2) e }$ and taking the  
limit as $\delta \rightarrow 2e$, we obtain,  
\be  
ds^2 = -\cG dT^2 + \cG dR^2 + \frac{\cG}{\cH} d\theta^2 +  
\frac{\cH}{\cG}\sin^2\theta \left( 2 a e R dT +   
\frac{e^2+a^2}{\chi^2}  
d \Phi \right)^2.  
\ee   
The electromagnetic field and potential again take the forms (\ref{UCF}) 
and
(\ref{UCA}).  
$R, T \in (-\infty, \infty)$, $\theta \in [0,\pi]$, and  
$\Phi$ inherits a period of $2 \pi$ from its predecessors.   
$\cG$, $\cH$, $X$, and $Y$ again retain their meanings from the Nariai 
case. 
 
Clearly the ($R$,$T$) sector of this space-time is conformally the same   
as two dimensional Minkowski flat space. There is no  
horizon structure, and therefore no black holes.

The physical parameters in both of these cases are given by  
\bea  
a^2 &=& \frac{3}{\Lambda} - 6e^2, \\  
M &=& 4\frac{\Lambda}{3} e^3, \mbox{ and}\\  
E_0^2 + G_0^2 &=& \Lambda e^4 + 6e^2 - \frac{3}{\Lambda},  
\eea  
and the allowed range of $e = \sqrt{\frac{3}{\Lambda}} X$ is given   
by,  
\bea  
\sqrt{-1 + \frac{2}{\sqrt{3}}} \leq  
& X & \leq \frac{1}{\sqrt{6}}.  
\eea  
  
Once more we note that when $a=0$ these
ultracold cases reduce to the two non-rotating ultra-cold solutions
considered in \cite{mann}. As noted above, neither of these
space-times contains black holes. Still for completeness, we shall
continue to include them in our considerations for the rest of the
paper. 
  
\subsection{Issues of Equilibrium}  
\label{equil}  
Before passing on to the next section where we will construct  
instantons to create the above space-times, we pause  
to examine whether these solutions to the Einstein-Maxwell  
equations are stable with respect to semi-classical effects. To  
this end, we must consider thermodynamically driven particle   
exchange  
between the horizons, electromagnetic discharge of the holes (due to  
emission of charged particles), and spin-down of the holes 
(due to emission of spinning particles and super-radiance).   
  
It is well known that a black hole emits particles in a black body   
thermal  
spectrum and thus may be viewed as having   
a definite temperature \cite{hawkNat}. In the same way, it has   
been shown that deSitter horizons may also be viewed as black bodies   
and have a definite temperature \cite{cosmo}. For a space-time with  
non-degenerate horizons, these  
temperatures may be most easily calculated by the conical singularity 
procedure  
\cite{OrigPathInt} (which we will return to when we construct   
instantons in the following sections). First, corotate the coordinate  
system with the horizon for which we are calculating the temperature.  
Second, analytically continue  
the time coordinate to imaginary values. For definiteness we will  
label the imaginary time  
coordinate $\cT$, the radial coordinate $\cR$, and let the horizon be  
located at $\cR = \cR_h$. Next,  
consider a curve in the $\cT-\cR$ plane with constant radial coordinate  
$\cR = \cR_0$.  
Periodically  
identify the imaginary time coordinate with some period $P_0$ so that   
this  
curve  becomes a coordinate ``circle'' and may be assigned a radius  
$R_0$ and circumference $C_0$ according to the integrals
\be  
R_0 \equiv \left. \left( \int_{\cR_h}^{\cR_0} \sqrt{g_{\cR \cR}} d\cR  
\right) \right|_{\cT=0},  
\;\;\;\;\;   
\mbox{and} \;\;\;\;\;   
C_0 \equiv \left. \left( \int_0^{P_0} \sqrt{g_{\cT \cT}} d\cT  
\right) \right|_{\cR=\cR_0}.  
\ee  
Finally, calculate $\lim_{\cR_0 \rightarrow \cR_h}   
\frac{C_0}{R_0}$.  
Pick the value of $P_0$ so that the limit has value $2\pi$. Then, the  
horizon has temperature $T_h = 1/P_0$, and surface gravity $\kappa_h =  
2\pi/P_0$. 
 
If there is a degenerate horizon, as is the case for a cold black hole,  
then that horizon is an infinite proper distance from all non-horizon   
points of  
the space-time. In such a situation there is no restriction on the period  
with which we may periodically identify the degenerate horizon, and it  
has been argued \cite{hhr} that the black hole can therefore be in  
equilibrium with thermal radiation of any temperature.   
  
We now consider which of our space-times are in thermodynamic   
equilibrium.  
First, consider the general, non-extreme KNdS solutions. The   
temperature  
of the outer black hole horizon and the cosmological horizon are  
respectively,   
\be  
T_{bh} = \left. \left( \frac{1}{4 \pi \chi^2 (r^2 + a^2)} \frac{d \cQ}{d  
r} \right) \right|_{r=r_{bh}} \;\;\; \mbox{and} \; \; \; T_{ch} = \left.  
\left(\frac{-1}{4 \pi \chi^2 (r^2 + a^2)} \frac{d \cQ}{d    
r} \right) \right|_{r=r_{ch}}  
\ee   
These two temperatures are equal if and only if, $4\lt \varepsilon^2 e  
(2e^2 - 2a^2 - \varepsilon^2 - \delta^2) = 0$. $\varepsilon=0$   
corresponds  
to the Nariai solutions which we will consider momentarily. $e=0$ is  
disallowed by the root ordering conditions (\ref{rootorder}). This   
leaves us  
with $2e^2 - 2a^2  
- \varepsilon^2 - \delta^2=0$ as the only case in which the non-  
extreme  
solutions achieve thermodynamic equilibrium. We label this the   
lukewarm  
case, in accordance with the instanton labeling scheme of   
\cite{mann}.     
We will consider the parameterization of these solutions in  
the next subsection.  
  
The cold limit is in thermodynamic equilibrium at the temperature of  
the cosmological horizon, for  
as we have noted an  
extreme black hole may be in equilibrium with thermal radiation of  
any temperature. The Nariai limit too is in thermodynamic   
equilibrium.  
Both the horizons have the same temperature,  
\be  
T_{Nar} = \frac{\lt (4e^2-\delta^2)}{4 \pi}.  
\ee   
The first ultracold case has only one horizon with temperature  
\be  
T_{UCI} = \frac{1}{2\pi},  
\ee  
and so with no other horizon to balance this one off, it is not in thermal  
equilibrium. The second ultracold case has no horizons, and so is trivially 
in 
equilibrium.
  
Next we consider discharge of the black holes. Even if the black hole   
and  
cosmological horizon are in equilibrium with respect to net particle  
exchange between them, there will be a  
net exchange of charge between the horizons. The mechanism is that   
even  
though the net numbers (and masses) of created particles may be the   
same,  
an excess of charged particles will be created at the black hole horizon,  
and so it will discharge \cite{discharge}. This effect may be   
completely    
avoided  
if there are no particles of the appropriate charge that are also lighter  
than the black hole. Thus, if magnetic monopoles do not exist   
then the magnetic holes will be stable with respect to  
discharge. Further, even if the appropriate light charged particles exist,  
the discharge effects will be small if the temperature of the black hole  
is small relative to the mass of those particles.  
  
Finally we consider the spin down of the black holes. If the black  
hole and cosmological horizons are at the same temperature, then there  
will be no net energy exchange between the horizons, but the particles  
created at the black hole horizon may still have an excess of angular  
momentum relative to those created at the cosmological horizon. This  
effect will tend to be small by itself, but it may be amplified by  
super-radiance. At this point things become somewhat complicated.   
Fairly  extensive investigations have been conducted into super-radiance   
effects in the asymptotically flat case \cite{page1,page2,chambers}, 
but only preliminary results are available for the  
asymptotically deSitter case \cite{maeda}. In particular, in   
\cite{maeda}  only one specific class of non-extreme holes have been
studied, and that class is not in thermodynamic equilibrium with 
the cosmological horizon. 

With these caveats in mind, consider the following.  
The original work by Page \cite{page1,page2} showed that  
for a wide class of massive and massless bosonic and fermionic fields  
(including the set of fields that we assume to exist in our universe), a  
spinning black hole (in an asymptotically flat universe) will radiate all  
of its angular momentum well before it has radiated all of its mass. He  
also speculated however, that given a large enough number of massless   
scalar  
fields, then the ratio of angular momentum to mass would approach a   
finite  
value rather than zero. Chambers, Hiscock, and Taylor showed that   
this in  
fact would be the case if 32 massless scalar fields exist   
\cite{chambers}.   
Maeda and Tachiwaza \cite{maeda} showed that a class of uncharged   
near-extremally  
rotating black holes in an asymptotically de Sitter space will spin down  
faster than their counterparts in asymptotically flat space-time. As   
noted  
before these holes were not in equilibrium with the cosmological   
horizon.   
  
Thus, within limits of current knowledge, it is consistent that  
the rotating holes will spin down to static black holes within finite  
time. What is not clear however is what the time scale for  
these effects is (particularly in the case where the horizons are in   
equilibrium). Further, the extant papers all agree with the physically  
intuitive  
idea that if the angular momentum is very small relative to the mass,   
then  
the rate of discharge of the holes will be small. These issues will be  
quantitatively investigated in a future paper.   For the remainder of this  
paper we shall consider only those situations for which discharge and   
spin-down  
effects may be neglected.  
  
\subsection{The Lukewarm Solution}  
\label{luke}  
  
As discussed in the previous subsection, the lukewarm solution is  
characterized by $2e^2 - 2a^2 - \varepsilon^2 - \delta^2=0$. We can   
use  
this relation to eliminate $\delta$ from the parameterizations of the  
physical  
parameters. We then have:  
\bea  
a^2 &=& 4e^2 - \frac{3}{\Lambda} \\  
M   &=& 2e(1 - \lt(3 e^2 + \varepsilon^2)) \\  
E_0^2 + G_0^2 &=& - \lt (7e^2+\e^2)(e^2-\e^2)-2(e^2-\e^2)+  
\frac{3}{\Lambda}.  
\eea  
Note that in this case, the expression for the charge may also be   
written  
as $E_0^2 + G_0^2 = \frac{M^2}{\chi^2} - a^2 \chi^2$.   
  
The range of the parameters is limited by the relations:  
\bea  
0 &\leq E <& 1 \\  
\frac{1}{\sqrt{5-2E-E^2}} &\leq X <& \sqrt{\frac{2}{E^2+7}}   
\label{lw2}\\   
\frac{1}{2} &\leq X \leq& \sqrt{ \frac{2 \sqrt{2-E^2} - 1 -  
E^2}{(E^2+7)(1-E^2)}} \label{lw3},  
\eea  
where as earlier $\e = E e$ and $e = \sqrt{\frac{\Lambda}{3}} X$.   
The  
second  
condition above is the $1 < \Delta < 2-E$ inequality for this case,   
while  
the third is the $a^2 \geq 0$, $E_0^2+G_0^2 \geq 0$ condition.   
Plotting  
the two conditions over the allowed range of $E$ we find that   
(\ref{lw2})  
is redundant, and so the lukewarm range is given by the first and third  
conditions.  
  
These space-times are non-extreme KNdS space-times, and so have the
global structure displayed in figure \ref{PenReg}. This space-time
was first discussed in \cite{MM}. Just as for the other special KNdS 
space-times that we considered in the absence
of rotation, the lukewarm case reduces to its non-rotating
counterpart discussed \cite{mann}.

\section{Instanton Construction}     
\label{InsCons}     
    
In this section we construct the instantons that will be used to study the     
creation rates of the space-times of the previous section. As discussed in     
the review of the path integral formalism, these instantons must both be     
solutions to the Einstein-Maxwell equations and also match smoothly onto     
the space-time that they are creating along a space-like hypersurface. The   
instantons that we construct will satisfy the cosmological no boundary  
condition,   
and so we will not need to worry about matching to initial conditions.  
     
\subsection{Step 1 - Analytic Continuation}     
    
In the construction of instantons for static spacetimes, the usual approach is  
to  analytically continue $t \rightarrow i t$. For a static space-time  
expressed in   
appropriate coordinates this gives a real Euclidean solution to the equations   
of motion but for a stationary space-time it produces a complex  
solution to the equations of motion. For now we accept this complex 
solution - later on in this section we will consider its relative merits 
compared to the more standard instanton where other metric parameters are 
also analytically continued in order to obtain a real Euclidean metric. We
proceed in the following manner  
(which is equivalent to continuing $t \rightarrow i \tau$). 
 
Foliating a space-time with a set of space-like hypersurfaces     
$\Sigma_t$ labelled by a time coordinate $t$ as in section \ref{actions}, 
we may in general write a Lorentzian metric as     
\bea     
\label{lapshif}     
ds^2 &=& -N^2 dt^2 + h_{ij}(dx^i + V^i dt)(dx^j + V^j dt) \\     
     &=& (-N^2 + h_{ij}V^iV^j) dt^2 + 2 h_{ij}V^j dx^i dt + h_{ij}dx^i     
dx^j, \nn     
\eea     
where $h_{ij}$ is the induced metric on the hypersurfaces, $N$ is the     
lapse function, and $V^i$ is the set of shift vector fields (a three     
vector field defined on each hypersurface).     
     
We now analytically continue using the prescription \cite{BMY} 
$N =  i \tN$ and $V^i = i \tV^i$, which is equivalent to analytically 
continuing $t\rightarrow i t$. The metric then becomes  
\be   
\label{lapshifInst}     
ds^2 = (\tN^2 - h_{ij}\tV^i\tV^j) dt^2 + i 2 h_{ij}\tV^j dx^i dt +     
h_{ij}dx^i     
dx^j.     
\ee   
If  $\tV^i=0$ then this metric has a Euclidean signature, whereas if 
$\tV^i \neq 0$ then  the   
metric is complex and its signature is not so easily defined. There is a sense   
however in which it is still Euclidean. At any point $x_0^\az$ we may  
make a  complex coordinate transformation $x^j = \tilde{x}^j - i\left. V^j  
\right|_{x=x_0} t$ (or equivalently add a complex constant to the shift),  
to obtain the metric 
\be     
\left. ds^2 \right|_{x^a_0} = \tN^2 dt^2 + h_{ij} dx^i dx^j,     
\ee     
at $x^a_0$. Thus the signature is Euclidean at any point modulo a  
complex coordinate transformation. For the electromagnetic field we set 
\be     
F_{tj} = i \tF_{tj}, \; \; F_{jt} = i \tF_{jt}, \mbox{     
and} F_{jk} = \tF_{jk}. 
\ee   
If the original Lorentzian metric and electromagnetic field were solutions 
to the Einstein-Maxwell equations, then so are this complex metric and 
electromagnetic field.

 We now proceed with the instanton construction by showing that these  
complex  solutions properly match onto their real counterparts.  
     
\subsection{Step 2 - Matching the instanton to the Lorentzian solution}     
 
The obvious hypersurface along which to match the Lorentzian solution to     
its complex ``Euclidean'' counterpart described above, is a  
$t=\mbox{constant}$     
hypersurface. We specialize the general metric (\ref{lapshif}) to the     
stationary, axisymmetric case where $x^1 = \phi$, $x^2 = \theta$, and $x^3     
= r$. Then, $V^i = [V^\phi(r,\theta), 0, 0]$, $N = N(r,\theta)$,     
and     
$h_{ij} = \mbox{diag} [h_{\phi \phi} (r,\theta), h_{\theta     
\theta}(r,\theta),     
h_{r r} (r,\theta)]$. We further restrict the electromagnetic field tensor     
such that $F_{t \phi}=F_{r \theta}=0$. This specialization will     
remain general enough to     
cover all of the cases in which we are interested.     
     
The normal vector to a $t=\mbox{constant}$ hypersurface $\Sigma_t$ in a     
Lorentzian solution is given     
by $u_\az = - N dt$. Then, with $e^\az_i$ being the projection     
operator taking     
vectors in $M$ on the $\Sigma_t$ into spatial vectors on $\Sigma_t$, we     
may calculate the crucial matching quantities $h_{ij}$, $K_{ij}$, $E_i$,     
and $B_i$ as follows:     
\be     
h_{ij} \equiv e_i^\az e_j^\bz (g_{\az \bz} + u_\az u_\bz) =  
\mbox{diag}[h_{\phi   
\phi},     
h_{\theta \theta}, h_{rr}],     
\ee     
\be     
K_{ij} \equiv e_i^\az e_j^\bz u_{\az;\bz} = \left[      
         \begin{array}{ccc}     
    0 & \frac{h_{\phi \phi} \partial_\theta V^\phi}{2N} &     
        \frac{h_{\phi \phi} \partial_r V^\phi}{2N} \\     
    \frac{h_{\phi \phi} \partial_\theta V^\phi}{2N} & 0 & 0 \\     
    \frac{h_{\phi \phi} \partial_r V^\phi}{2N} & 0 & 0     
         \end{array}     
         \right],     
\ee     
\be     
E_i \equiv e_i^\az F_{\az \bz} u{^\bz} = \left[0 , \frac{F_{\theta t}     
- F_{\theta \phi} V^\phi}{N}, \frac{F_{r t} - F_{r \phi}V^\phi}{N}\right],     
\mbox{ and}     
\ee     
\be     
B_i \equiv -\frac{1}{2} e_i^\az g_{\az \bz} \varepsilon^{\bz \cz \dz  
\ez}u_\cz   
F_{\dz \ez}     
= \left[0, -\frac{h_{\theta \theta} F_{\phi r}}{\sqrt{h_{\phi     
\phi}h_{\theta     
\theta} h_{rr}}}, \frac{h_{rr} F_{\phi \theta}}{\sqrt{h_{\phi     
\phi}h_{\theta \theta} h_{rr}}} \right].     
\ee      
Switching to the complex ``Euclidean" solution via the preceding  
prescription 
we see that the four surface quantities $h_{ij}$, $K_{ij}$, $E_i$, and  
$B_i$  
are invariant under this set of transformations and so we can smoothly  
match the     
Euclidean and Lorentzian solutions along a $t=\mbox{constant}$     
hypersurface.   
     
Before passing on to consider the instantons that may be constructed from  
these  complex solutions, we note that in dealing with complex solutions  
we have  
made  a departure from the usual method of instanton construction used in   
\cite{MM,LP,wu}. The standard method would require that we analytically  
continue as many parameters of the metric as necessary so that we would  
arrive at a  
real and Euclidean solution to the Einstein-Maxwell equations. For  
example,  with the KNdS solutions we would continue $a \rightarrow  
i\tilde{a}$   
which would make the metric real and Euclidean and $E_0 \rightarrow    
i\tilde{E_0}$ so that the electromagnetic field would be real.  
Although this approach avoids dealing with complex metrics, it  
incurs   
several serious problems of its own. Specifically, if we complexify $a$ and   
$E_0$ then the structure of many components of the KNdS metric will   
change; for example $\cQ \rightarrow -\frac{\Lambda}{3} r^4  
+ (1 + \frac{\Lambda}{3} \tilde{a}^2)r^2 - 2 M r - \tilde{E_0}^2 +  
G_0^2$.  
 
Such a change will alter the root structure of $\cQ$. Depending on the 
relative magnitudes of the parameters, the number of roots of $\cQ$ can  
change and 
the roots corresponding to the cosmological and outer black hole horizons  
can vanish. If the root structure of the Lorentzian solution  
does not match that of its ``Euclidean''  counterpart 
then clearly we cannot match them along a spatial hypersurface.  
Even if  the number of roots remains constant, the change in $\cQ$ (as  
well as  $\cG$, $\cH$, and $\chi^2$), will mean that the induced metrics,  
extrinsic   
curvatures, and electric and magnetic fields on $\Sigma$ will no longer  
match.  Thus, such a Euclidean solution will not match onto the real  
Lorentzian  
solution according to the standard prescription, and we cannot  
demand both that the instanton be real and that it match the Lorentzian  
solution  along a $t = \mbox{constant}$ hypersurface. Given that the  
matching conditions are the only conditions available that 
prescribe the connection between 
the instantons and the physical Lorentzian solutions 
we choose to keep the matching conditions and abandon the requirement that
the metric be real.
     
\subsection{Putting the Parts Together}     
\label{ppt}

We are now ready to finish off the instantons.     
They will come in three classes: i) those creating spacetimes with two     
non-degenerate horizons bounding the primary Lorentzian sector (this case     
will create Nariai and lukewarm spacetimes), ii) those creating      
spacetimes with only a single non-degenerate horizon bounding the     
Lorentzian sector,      
(this case will create cold spacetimes and ultracold I spacetimes),     
and iii) and those creating zero horizon spacetimes (here, the ultracold II     
space-time).     
     
\subsubsection{Spacetimes with two nondegenerate horizons}     
By the procedure described above we have constructed a complex     
solution that may be joined to the Lorentzian solution from which it was     
generated. However a subtlety arises in that the     
$t=\mbox{constant}$ spatial hypersurfaces of the nondegenerate KNdS  
and     
Nariai spacetimes both consist of two Lorentzian regions that are     
connected to each other across their corresponding horizons, while the     
$t=\mbox{constant}$ hypersurfaces of the complex solution consist of  
only one   
such region. The complex solution may be connected to both sections     
simultaneously by the following procedure (that is also illustrated in     
figure \ref{H2hor}).     
     
\begin{figure}     
\centerline{\psfig{figure=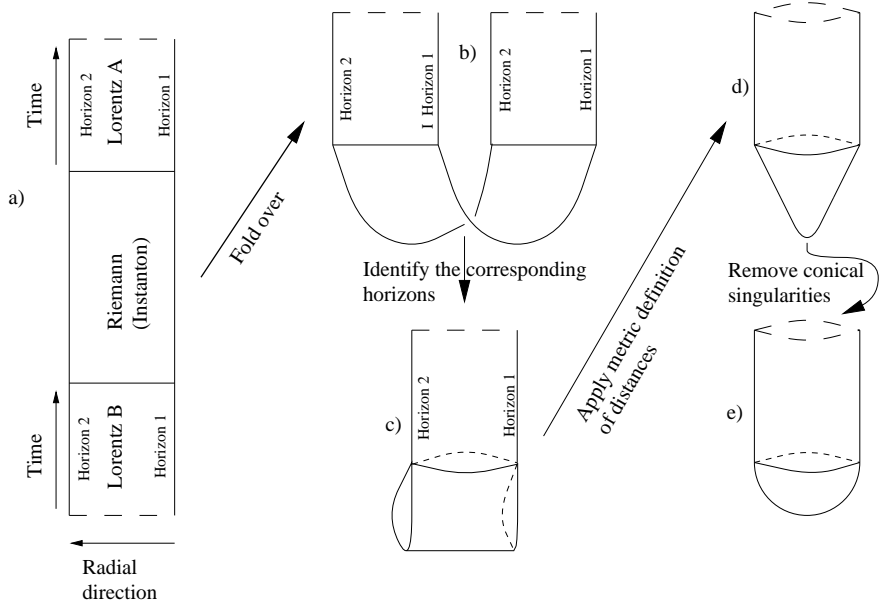,height=8cm,angle=270}}     
\caption{Construction of the two horizon instanton. The radial/time     
sector is shown. The heavily dashed lines indicate that the solution     
continues in that direction.}     
\label{H2hor}     
\end{figure}     
     
First, at two $t = \mbox{constant}$ ($t=0$ and $t=\frac{P_0}{2}$)  
hypersurfaces  
connect half of a full Lorentzian solution (a region bounded by the  
outer black hole and cosmological horizons) 
as in figure \ref{H2hor}a. Next (figure \ref{H2hor}b) fold the construction  
over,  
and identify outer horizon to outer horizon, and inner horizon to inner  
horizon  
(figure \ref{H2hor}c). The $t=\mbox{constant}$ hypersurfaces of the  
Lorentzian part of     
the construction now consist of two regions with opposite spin and charge,     
and are the complete $t=\mbox{constant}$ hypersurfaces of the maximally 
extended and periodically identified KNdS solutions that we considered 
earlier.
     
Next note that the metric at any point of the Riemannian part of the     
construction is      
\be     
ds^2 = \tilde{N}^2 dt^2 + h_{ij}dx^i dx^j 
\ee     
under the coordinate transformation that eliminates the shift at that point. 
At the horizons $\tilde{N}^2 \rightarrow \frac{\cQ \cG}{(r^2+a^2)^2
\chi^4} \rightarrow 0$     
for the lukewarm (Nariai) solutions. Therefore it is reasonable to     
identify the entire time coordinate along the horizons as a single time     
(figure \ref{H2hor}d). The instanton is nearly complete.  
The Riemannian part is smooth everywhere except possibly where  
we have made the identification along the horizons where the procedure 
may induce conical singularities, in violation of the Einstein equations.  
 
For a given horizon at $r=r_h$, if we choose $P_0$ such that  $\lim_{r     
\rightarrow r_h} \frac{ P_0 \partial_r \tilde{N}}{\sqrt{h_{rr}}} = 2\pi$
then the conical singularity is eliminated at that point. This 
is the same condition used in calculating the temperature of the     
horizons in section \ref{equil}, and so we may simply apply our results     
from there. Hence the only double-horizon cases where the conical 
singularities at the two horizons may be simultaneously eliminated  (figure 
\ref{H2hor} e)
-- implying that the instanton will everywhere be a solution to     
the Einstein equations -- will be the lukewarm and Nariai instantons,     
for which     
\be     
P_0^{lw} = \frac{4 \pi \chi^2 (r_{bh}^2 + a^2)}{Q'(r_{h})} \ \mbox{and} 
\ \     
P_0^{Nar} = \frac{4 \pi}{\frac{\Lambda}{3} (4e^2 - \delta^2)},     
\ee     
where $Q' = \frac{d Q}{d r}$, and $r_{bh}$ is the radius of the outer black  
hole horizon in the luke warm solution. We next consider the single- 
horizon spacetimes.

\subsubsection{Spacetimes with one non-degenerate horizon}     
 
With the double-horizon instanton construction completed, the single     
non-degenerate horizon instantons come more easily. These are the cold  
and   ultracold I  spacetimes. Note that even though the cold  
space-time has two horizons,       
the inner horizon is a degenerate, double horizon.     
For these spacetimes, we still attach     
half-copies of the Lorentzian space-time at the $t=0$ and     
$t=\frac{P_0}{2}$ hypersurfaces of the complex Riemannian section  
(figure      
\ref{H1hor}a).     
Then we fold and identify the cosmological horizons (thus reconstructing     
the full Lorentzian $t=\mbox{constant}$ hypersurfaces (figure  
\ref{H1hor}b,c).     
Next, we again identify the time coordinate along the cosmological horizon     
(figure \ref{H1hor}d). Finally, with just one horizon we choose     
\be     
P_0^{cold} = -\frac{4 \pi \chi^2 (r_{ch}^2 + a^2)}{Q'(r_{ch})}     
\qquad \mbox{ and} \qquad     
P_0^{UCII} = 2\pi,     
\ee     
where $Q'(r_{ch}) = \left. \frac{d Q}{d r} \right|_{r=r_{ch}}$, $r_{ch}$ is  
the radius of the cosmological horizon. Then the instanton will have no  
conical singularities     
(figure \ref{H1hor}e). 
 
\begin{figure}     
\centerline{\psfig{figure=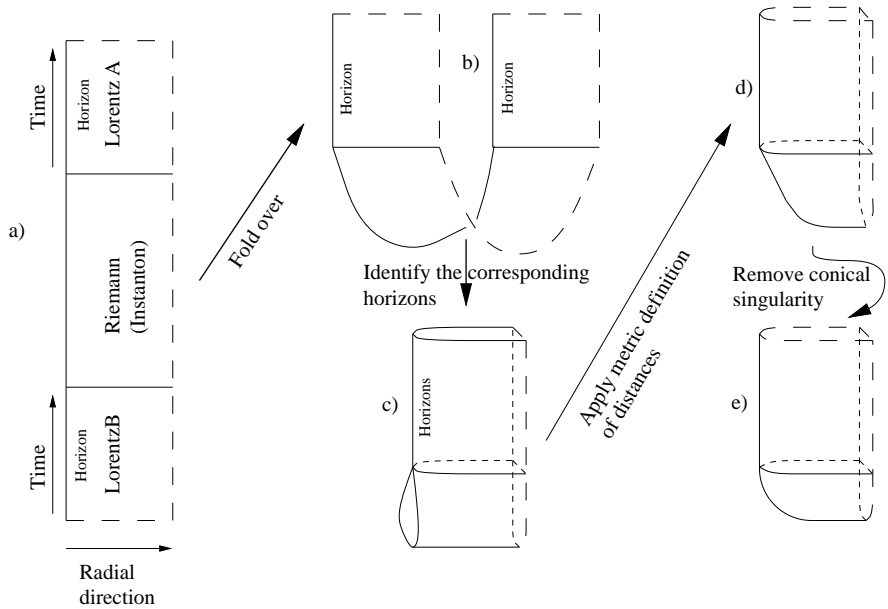,height=8cm,angle=270}}     
\caption{Construction of the one horizon instanton. The radial/time      
sector is shown. The heavily dashed lines indicate that the solution      
continues in that direction.}     
\label{H1hor}     
\end{figure}

\subsubsection{No-horizon spacetimes}

This time the construction is less definite. With no identifications being     
made, and no horizons to define a period, we have an instanton of     
indefinite period creating two disjoint spacetimes (figure \ref{H0hor}).     
This corresponds to the ultracold II case.

\begin{figure}     
\centerline{\psfig{figure=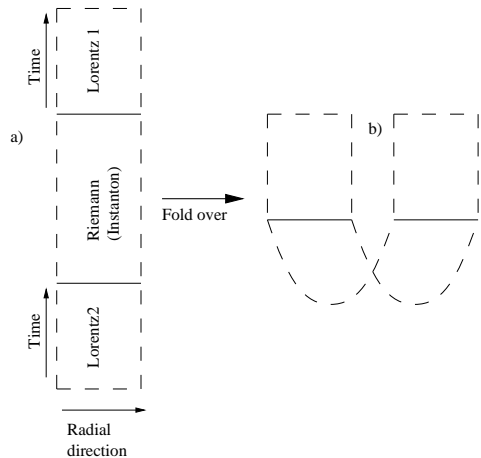,height=6cm,angle=270}}     
\caption{Construction of the no horizon instanton. The radial/time      
sector is shown. The heavily dashed lines indicate that the solution      
continues in that direction.}        
\label{H0hor}     
\end{figure}

\section{Action Calculation}    
\label{actions}    
    
In this section we shall calculate the actions of the instantons of the    
previous section. From these actions we may then estimate the creation    
rates and entropy for the associated spacetimes. However, in order to do    
these calculations properly we must ensure that we are using the correct    
form of the action. To this end, we quickly review and then apply the     
quasilocal formalism of Brown and York \cite{BY}. Since we are    
interested    
in KNdS black holes, we shall include electromagnetic fields in the    
calculations.     
    
\subsection{Idea and Definitions}    
In order to extract the physics of the Einstein-Maxwell action, the    
quasi-local formalism analyses a finite region of space-time using a    
Hamiltonian approach. That is, the region is foliated by a set of    
space-like hypersurfaces which intuitively are surfaces of simultaneity    
representing ``instants'' of time. A flow is then defined over the region    
to describe the passage of time from instant to instant. With these two    
concepts in place, the action may be understood in terms of our    
intuitive concepts of energy, angular momentum, and    
stress-energy. In the following we apply this programme to our    
situation. Throughout the concepts will be illustrated by figure    
\ref{splog}.    
    
\begin{figure}    
\centerline{\psfig{figure=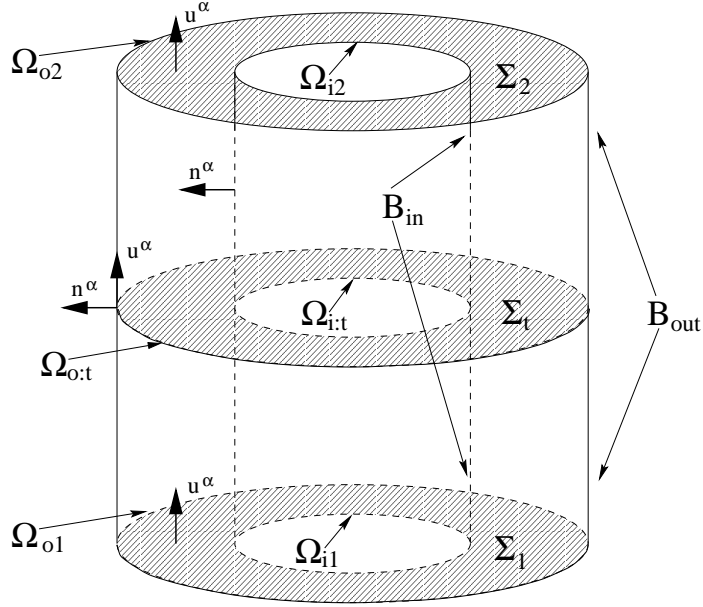,height=8cm,angle=270}}      
\caption{A three dimensional schematic showing the region $M$, its    
assorted boundary components, and the normals to those components.}    
\label{splog}    
\end{figure}    
    
We define a region of space-time $M$ as follows. Let $\cM$ be a    
space-time and let $M \subset \cM$ such that $\partial M$ consists    
of    
two space-like three surfaces $\Sigma_1$ and $\Sigma_2$ and two time-   
like     
three surfaces $B_{in}$ and $B_{out}$. These surfaces intersect as    
follows: $B_{out} \cap \Sigma_1 = \Omega_{o1}$, $B_{in} \cap    
\Sigma_1    
= \Omega_{i1}$, $B_{out} \cap \Sigma_2 = \Omega_{o2}$,    
$B_{in}    
\cap \Sigma_2 = \Omega_{i2}$, $B_{in} \cap B_{out} =    
\emptyset$,    
and    
$\Sigma_1 \cap \Sigma_2 = \emptyset$.     
    
Tensors in this space-time shall be labelled with greek indices. The metric    
tensor will be $g_{\alpha \beta}$ and the covariant derivative compatible    
with $g_{\alpha \beta}$ will be $\nabla_\alpha$. The Riemann tensor,    
Ricci tensor, Ricci scalar, and Einstein tensor will be $\cR_{\alpha \beta    
\delta \gamma}$, $\cR_{\alpha    
\beta}$, $\cR$, and $G_{\alpha \beta}$ respectively.    
    
Consider the time-like surfaces $B_{in}$ and $B_{out}$, starting with    
$B_{out}$. We define $n^\alpha$ to be the outward pointing    
space-like unit    
normal vector field on $B_{out}$. Then, we may define    
$\gamma^\alpha_{\;    
\; \beta} = \delta^\alpha_{\; \; \beta} - n^\alpha n_\beta$ as the    
projection tensor for $B_{out}$\footnote{It is a projection operator in    
the sense that that $\gamma^\alpha_{\; \; \delta} \gamma^\delta_{\;    
\; \beta} = \gamma^\alpha_{\; \; \beta}$ and if $p \in B_{out}$, and    
$v^\alpha \in T_p \cM$, then $\gamma^\alpha_{\;    
\; \beta} v^\beta \in T_p B_{out}$. Similarly, if $v_\alpha \in T^*_p    
\cM$, then $\gamma^\beta_{\; \; \alpha} v_\beta \in T^*_p B_{out}$ (the    
extension to more general tensor fields is made in the obvious way).}.     
Projecting the metric tensor onto $B_{out}$ we get the induced    
Lorentzian metric    
$\gamma_{\alpha \beta} = g_{\alpha \beta} - n_{\alpha} n_{\beta}$. The    
extrinsic curvature of this hypersurface in $\cM$ is then defined as    
$\Theta_{\alpha \beta} = - \gamma^\cz_{\;    
\; \az} \nabla_\cz n_\bz$. The trace of $\Theta_{\alpha    
\beta}$ is $\Theta =    
g^{\alpha \beta}    
\Theta_{\alpha \beta} = \gamma^{\alpha \beta} \Theta_{\alpha \beta}$.     
The    
same notation will also be used to denote the corresponding quantities on    
$B_{in}$ (though in that case the normal vector field will point into    
$M$).    
    
Next we consider the space-like surfaces $\Sigma_1$ and $\Sigma_2$    
starting    
with $\Sigma_2$. We define $u^\alpha$ to be the outward pointing    
time-like     
unit normal vector field on $\Sigma_2$. Then the projection tensor is     
$h^\alpha_{\; \beta} = \delta^\alpha_{\; \beta} + u^\alpha    
u_\beta$ and the induced Euclidean    
signature metric on the surface is $h_{\alpha \beta} = h^\delta_{\;    
\alpha} h^\epsilon_{\;    
\beta} g_{\delta \epsilon} = g_{\alpha \beta} + u_\alpha    
u_\beta$. The compatible covariant derivative is     
$D_\alpha$\footnote{Note that    
hypersurface covariant derivatives may be calculated by taking the    
covariant derivative of a tensor with    
$\nabla_\alpha$ and then projecting the result down into the hypersurface.    
For    
example if $A^\alpha_{\; \; \beta} \in T^1_1 \Sigma_2$, then $D_\gamma    
A^\alpha_{\; \; \beta} = h^\lambda_{\; \; \gamma} h^\alpha_{\; \; \mu}    
h^\nu_{\; \; \beta} \nabla_\lambda A^\mu_{\; \; \nu}$.}.     
The extrinsic    
curvature of $\Sigma_2$ in $\cM$ is $K_{\alpha \beta} = - h^\gamma_{\;    
\;    
\alpha} \nabla_\gamma u_\beta$, and its trace is $K= g^{\alpha \beta}    
K_{\alpha \beta} = h^{\alpha \beta} K_{\alpha \beta}$. The    
intrinsic Riemann tensor, Ricci tensor and scalar of the surface are    
$R_{\alpha \beta    
\gamma \delta}$, $R_{\alpha \beta}$, and $R$ respectively. Often, we    
shall    
tag tensors defined intrinsically on $\Sigma_2$    
with lower case mid-alphabet indices (eg.\ $h_{ij}$, $K_{ij}$). Then the    
compatible covariant derivative will be $D_i$. The same notations will    
also    
be used to denote the corresponding quantities on $\Sigma_1$ (though in    
this case the unit normal vector will point into $M$).    
      
Finally, we treat the intersection surfaces $\Omega_{o1}$,    
$\Omega_{o2}$, $\Omega_{i1}$, and $\Omega_{i2}$, starting    
with $\Omega_{o2}$. With loss of generality we shall make the standard    
assumption that on the $\Omega$ surfaces, $n^\az \bot u^\az$ (the    
non-orthogonal cases are partially discussed in    
\cite{hunter} and \cite{hayward} and will be    
discussed in more detail in \cite{me}). Then, the projection operator    
onto the $\Omega$ surface is $\sigma^\az_{\; \; \bz} = \delta^\az_{\; \;    
\bz} +    
u^\az u_\bz - n^\az n_\bz$, and the induced metric is $\sigma_{\az \bz} =    
\sigma^\az_{\; \; \cz} \sigma^\bz_{\; \; \dz} g_{\cz \dz} = g_{\az \bz} +    
u_\az u_\bz - n_\az n_\bz$. If $\Omega_{o2}$ is viewed as a surface    
embedded    
in    
$\Sigma_2$ then it has an extrinsic curvature of $k_{\alpha \beta} \equiv    
-\sigma^\az_{\; \; \cz} \sigma^\bz_{\; \; \dz} D_{\cz} n_\dz $ with    
respect to $\Sigma_2$. The trace of $k_{\az \bz}$ is $k = g^{\az \bz}    
k_{\az    
\bz} = \sigma^{ij} k_{ij}$. The same notations will be used to denote       
the corresponding quantities on $\Omega_{o1}$,    
$\Omega_{i1}$, and $\Omega_{i2}$.    
    
The next step of the programme is to define the surfaces of simultaneity -    
the ``instants'' of time.     
We decompose $M$ into a set of space-like hypersurfaces    
$\left\{    
\Sigma_t : t_1 \leq t \leq t_1 \right\}$ such that $\Sigma_{t_1} =    
\Sigma_1$ and $\Sigma_{t_2} = \Sigma_2$. On these surfaces we define    
the    
same quantities that were defined on $\Sigma_1$ and $\Sigma_2$. The    
same    
notation will be used for these quantities. The unit normals $u^\alpha$    
are chosen to be consistently oriented with those on the boundary    
surfaces and we label the intersections with $B_{in}$ and    
$B_{out}$ as $\Omega_{o:t} = B_{out}    
\cap    
\Sigma_t$, $\Omega_{i:t} = B_{in} \cap \Sigma_t$. We extend our    
earlier orthogonality assumption so that on these intersection surfaces    
$u^\az n_\az = 0$.    
    
Finally, we define a flow of coordinate time on $M$.    
If $t$ is the hypersurface label, then we choose a vector field    
$t^\alpha$ such that $t^\az \nabla_\az t = 1$. This vector field defines     
a flow. It may be decomposed into    
parts perpendicular and parallel to the hypersurfaces as    
\be    
t^\alpha = N u^\alpha + V^\alpha,    
\ee    
where $N$ is called the lapse function and $V^\alpha$ is the shift vector    
which lies in the hypersurfaces (that is $u_\alpha V^\alpha = 0$).    
Then the metric on $M$ may be decomposed 
\bea    
ds^2 = -N^2 dt^2 + h_{ij} \left( dx^i + V^i dt \right)    
\left( dx^j + V^j dt \right),    
\eea    
as in section \ref{InsCons}. It is then easy to show that  
 
$\sqrt{-g} = N \sqrt{h}$, and $\sqrt{-\gamma} = N \sqrt{\sigma}$.

\subsection{Analyzing the Action}    
    
The usual Einstein-Maxwell action with its boundary terms is,    
\bea    
I = -\frac{1}{2 \kappa} \int_M d^4 x \sqrt{-g} \left( \cR - 2\Lambda -    
F^2\right)    
+ \frac{1}{\kappa} \int_{\Sigma} d^3x \sqrt{h} K    
-  \frac{1}{\kappa} \int_{B} d^3x \sqrt{-\gamma} \Theta.    
\label{basicAction}    
\eea    
We will work in the coordinate system where $c = \hbar = 1$, and so    
$\kappa = 8 \pi$.     
Throughout this section an integral with subscript $B$ denotes 
two integrals - the indicated integral taken over $B_{out}$ minus    
the same integral over $B_{in}$. In the same way $\int_\Sigma =    
\int_{\Sigma_2} - \int_{\Sigma_1}$, $\int_\Omega =       
\int_{\Omega_{o2}} + \int_{\Omega_{i1}} - \int_{\Omega_{i2}} -    
\int_{\Omega_{o1}}$, and $\int_{\Omega_t} = \int_{\Omega_{o:t}} -    
\int_{\Omega_{i:t}}$.    
    
Decomposing the action according to the foliation and time    
flow yields \cite{BY,hawkinghorowitz} 
\bea    
\label{decomp} 
I &=& -\int_M d^4 x \left( P^{ij} \pounds_t h_{ij} - N \cH - V^i    
\cH_i - \frac{\sqrt{h}}{4 \pi} E^i \pounds_t A_i +    
\frac{\sqrt{h}}{4\pi} A_\alpha t^\alpha    
\cF_{el} \right)\\    
&& + \int dt \int_{\Omega_t} d^2 x \sqrt{\sigma} \left( N (\epsilon^{GR}     
+ \epsilon^{EM}) - V^i ( j_i^{GR} + j_i^{EM} ) \right), \nn    
\eea    
where $P^{ij} = \frac{\sqrt{h}}{16 \pi} \left( K h^{ij} - K^{ij}    
\right)$, $\pounds_t$ is the Lie derivative in the $t^\alpha$    
direction, $\cH$ and $\cH_a$ are the Einstein-Maxwell constraints    
(\ref{MEC1}) and (\ref{MEC2}),    
and    
$\cF_{el}$ is the electric Maxwell constraint (\ref{MC1}). 
$A^\alpha$ is the electromagnetic vector    
potential, and may be broken up into its components perpendicular to and    
parallel     
to the hypersurfaces as $A^\alpha \equiv -\Phi    
u^\alpha + \tilde{A}^\alpha$. $\varepsilon^{GR}$ and    
$\varepsilon^{EM}$    
are the    
energy densities of the gravitational and electromagnetic fields, while    
$j^{GR}_a$ and $j^{EM}_a$ are the angular momentum densities.    
Explicitly    
they are    
\bea 
\varepsilon^{GR} &=& \frac{k}{\kappa}, \\ 
\varepsilon^{EM} &=& \frac{2 \Phi E_i n^i}{\kappa}, \\ 
j_i^{GR} &=& -\frac{2 \sigma_{ij}n_k P^{jk}}{\sqrt{h}}, \mbox{ and}    
\\     
j_i^{EM} &=& \frac{2 \tilde{A}_i E_j n^j}{\kappa},    
\eea 
where $E_i = e^\alpha_i F_{\alpha \beta} u^\beta$ is the electric field  
induced on the hypersurfaces. 
The interpretation of these quantities as energy and angular momentum    
densities is supported by calculations in the Schwarzschild and Kerr    
spacetimes \cite{BY,hawkinghorowitz}.

We may calculate the variation of $I$ with respect to  
$g_{\alpha \beta}$ (equivalently $N$, $V^\alpha$ and    
$h_{\alpha \beta}$), and $A_\alpha$ (equivalently $\Phi$ and    
$\tilde{A}_a)$, as \cite{BY,me}: 
\bea    
\label{variation}    
\delta I &=&- \frac{1}{2 \kappa} \int_M d^4 x \sqrt{-g} \left\{ [G_{\az    
\bz}     
+    
\Lambda g_{\az \bz}  - 8 \pi T^{EM}_{\az \bz} ] \delta g^{\az \bz} - 4    
[\na_\az F^{\az \bz} ] \dz A_\bz \right\}  \\    
         && - \int_\Sigma d^3 x \left\{ P^{ij} \dz h_{ij} -    
\frac{2}{\kappa}    
\sqrt{h} E^i \dz \tilde{A}_i \right\} \nn \\    
         && - \int dt  \int_{\Omega_t} d^2 x\sqrt{\sigma} \left\{    
[\varepsilon^{GR} + \varepsilon^{EM}] \dz N - [j^{GR}_i + j^{EM}_i]    
\dz    
V^{i} - \frac{N}{2} s^{ij} \dz \sigma_{ij} \right\} \nn \\    
         && - \frac{1}{\kappa} \int dt  \int_{\Omega_t} d^2 x N    
\sqrt{\sigma}    
\left\{ (n^i    
F_{ij} \sigma^{jk}) \dz ( \sigma^l_{\; \; k} \tilde{A}_l ) +    
(n^i E_i) \delta \Phi \right\}. \nn    
\eea    
In the above $T^{EM}_{\alpha \beta}$ is the standard electromagnetic  
stress energy tensor while $s^{ij}$ is the stress tensor for the  
surfaces $\Omega_{i/o:t}$. This tensor may written in terms of a trace-free  
shear $\eta^{ij}$ and pressure $p$ as 
\be 
s^{ij} =  \frac{p}{2} \sigma^{ij} + \eta^{ij }, 
\ee  
while in turn the pressure and shear may be written as, 
\bea    
\label{pressure}    
p &=& \frac{1}{\kappa} \left( 2 \frac{n^i    
\partial_i N}{N} - k \right), \mbox{ and}\\    
\eta^{ij} &=& \frac{1}{\kappa} \left(k^{ij} - \frac{k}{2}    
\sigma^{ij}    
\right).    
\eea     
 
The last boundary term of $\delta I$ is purely electromagnetic and its  
components may be given a simple physical interpretation. To see this,  
recall that the electric and magnetic fields induced on the $\Sigma$  
surfaces are 
\bea    
\label{EMfromA}    
E_\az &\equiv& F_{\az \bz}u^\bz = \frac{1}{N} D_\az (N \Phi) -
\pounds_u {\tilde{A}_\az},    
\mbox{ and} \\    
B^\az &\equiv& - \frac{1}{2} \varepsilon^{\az \bz \cz \dz} u_\bz F_{\cz     
\dz} = - \varepsilon^{\az \bz \cz \dz} u_\bz D_\cz \tilde{A}_\dz, \nn    
\eea   
with respect to the field tensor $F_{\az \bz}$ and vector potential    
$A_\az$. Then we can see that $\sigma^{ij}\tilde{A}_j$     
and $\Phi$ are necessary and sufficient to fix the component of $B_i$  
perpendicular to the surfaces $\Omega_t$ and the components of $E_i$  
parallel to those same surfaces. By contrast, $n^i F_{ij} \sigma^{jk}$ and  
$n^i E_i$ are necessary and sufficient to fix the perpendicular component  
of $E_i$ and the parallel components of  $B_j$. A simple application of  
Gauss's law of electromagnetism to the boundaries $\Omega_t$ of the  
regions $\Sigma_t$, then reveals that $\sigma^{ij}\tilde{A}_j$     
and $\Phi$ are sufficient to fix the magnetic (but not electric) charge  
contained in the hypersurfaces $\Sigma_t$ while  $n^i F_{ij} \sigma^{jk}$  
and $n^i E_i$ are sufficient to fix the electric (but not magnetic) charge  
contained in the hypersurfaces. 
 
We may find extremal points of the action functional by setting $\delta I
= 0$. Then, the two bulk terms of $\delta I$ respectively give us the
Einstein-Maxwell equations. The remaining
boundary terms specify quantities that must be fixed when we consider this
particular action functional. Thus, in this case, the induced
metric $h_{ij}$ and
$\tilde{A}_j$ (and therefore the magnetic field) are fixed on the
$\Sigma_1$ and $\Sigma_2$  surfaces, while the lapse $N$, shift $V^i$,
induced metric
$\sigma_{ij}$, $\sigma^i_{\ j} \tilde{A}_i$, and $\Phi$ are fixed   
on the boundaries $\Omega_t$. By the discussion of the previous 
paragraph,
this means that we are considering paths with a fixed magnetic charge 
when
we use this action functional.

We could change this situation if we chose to use the action
functional $I_{electric} \equiv    
I + \Delta I_{electric}$, where    
\bea
\label{deltaElectric}   
\Delta I_{electric} \equiv -\frac{1}{\kappa} \int_\Sigma d^3 x \sqrt{h}    
F^{\alpha \beta} u_\alpha A_\beta + \frac{1}{\kappa} \int_B d^3 x    
\sqrt{-\gamma} F^{\alpha \beta} n_\alpha A_\beta,    
\eea
instead. Solving $\delta I_{electric} = 0 $ we obtain the same
equations of motion, but this time the purely electromagnetic boundary
terms become
\bea    
- \frac{2}{\kappa} \int_\Sigma d^3 x  
\sqrt{h}  \tilde{A}_i \dz E^i + 
\int dt \int_{\Omega_t} N \sqrt{\sigma} \left\{ ( \sigma^{\bz \cz}    
\tilde{A}_\cz ) \dz (n^\az F_{\az \bz} \sigma^{\bz \cz}) +    
\Phi \delta(n^a E_a) \right\}.    
\eea
Then, this modified action functional fixes
the electric field on $\Sigma_1$ and $\Sigma_2$ and 
by equations (\ref{EMfromA})  the electric (but not magnetic) charge on 
the $\Sigma_t$ hypersurfaces.

In a similar way we could (and in fact will) choose to fix the angular
momentum of the paths considered by adding
\bea    
\label{angmomterm}    
\Delta I_{AngMom} &=& - \int dt \int_{\Omega_t}  d^2 x \sqrt{\sigma}    
V^i    
(j^{GR}_i + j^{EM}_i )    
\eea    
to $I$. Then $(j^{GR}_i + j^{EM}_i)$ rather than $V^i$ will be fixed on  
the $\Omega_t$ boundary surfaces. Hence the $\Sigma_t$ hypersufaces 
will all have the same total angular momentum. Thus with this 
modification to the action, paths contributing to the path integral will have 
fixed angular momentum.

\subsection{Choosing an Action}    
    
Hence a choice of action entails a choice of boundary    
conditions that must be satisfied by solutions to the equations of motion.  
These   boundary conditions are crucial to a correct application of the path    
integral formulation of gravity.    
A path integral with final conditions $X_2$ may be interpreted as a    
sum over all    
possible histories of a system leading up to the state $X_2$. Given this    
interpretation, and assuming that the state $X_2$ is in thermodynamic    
equilibrium, we may reinterpret the path integral as a thermodynamic    
partition function. Then the boundary conditions chosen along with an    
action become restrictions on which histories contribute to the    
partition function. The application of these restrictions then defines    
exactly which partition function we are studying -- {\it i.e.} whether it is    
canonical, microcanonical, grand canonical, or some less standard    
partition function.    
    
Pair creation calculations are typically carried out in the 
canonical partition    
function -- that is where the temperature and all     
extensive variables (angular momentum,     
electric and/or magnetic charge) except for the energy are fixed    
\cite{hawkingross}.    
This ensures that    
created spacetimes are in thermal equilibrium, that 
there is no discontinuity in physical properties such as
electromagnetic charge 
and angular momenta at the juncture of the paths and the Lorentzian
solution, and
from a geometric point of view that the paths will smoothly match onto
the Lorentzian solution. At first it might seem unusual that we do not
choose to fix the energy, but support for    
this choice of partition function may be found if we examine the boundary    
conditions which must be imposed such that our interpolations will be    
smooth at the horizons.      
    
At a non-degenerate horizon ({\it i.e.} both horizons of the lukewarm and    
Nariai instantons, the cosmological horizon for the cold instanton, and    
the single horizon for the ultracold I instanton) the paths will be closed    
and smooth if and only if $\tilde{N}=0$ and there are no conical
singularities    
at those horizons. These conditions were discussed in some  
detail in    
section \ref{ppt} for the actual instantons, and may be extended without    
difficulty to general paths. Recall that a conical    
singularity exists at a non-degenerate horizon at $r_h$ unless
\bea    
\lim_{r \rightarrow r_h} \frac{\int_0^{P_0/2} dt \tilde{N}}{\int_{r_h}^r    
dr    
\sqrt{h_{rr}}} = \pi    
\eea    
This is equivalent to the condition 
\bea   
\label{periodfix}    
\lim_{r\rightarrow r_h} \frac{P_0 \partial r \tilde{N}}{2 \sqrt{h_{rr}}} =      
\pi    
\eea    
and so to ensure that all paths smoothly close at the horizons, this
quantity must be fixed there. Since $\tilde{N}$ already vanishes at    
the  horizon and since $n^i = \frac{1}{\sqrt{h_{rr}}}    
\frac{\partial}{\partial r}$, we see that this is exactly equivalent to    
fixing $\tilde{N} p$, where $p$ is the pressure defined in    
(\ref{pressure}). Hence wherever we have a non-degenerate horizon 
we must add a boundary term 
\bea    
\Delta I_{pressure} = -\frac{1}{2} \int dt \int_{\Omega_{r_h:t}} d^2 x    
\sqrt{\sigma} \tilde{N} p,    
\eea    
at that horizon
in order for all paths to be smooth at those points. This has the effect of 
fixing the temperature at the horizon since (see section \ref{equil})  
constraining an interpolation to be regular at a horizon is equivalent to  
fixing the temperature there. Thus, for     
the two non-degenerate horizon spacetimes, geometric regularity demands
that we fix the lapse $\tilde{N}$ and pressure $p$ (equivalently the 
temperature)
rather than the energy densities.

Consider next the cold case with a degenerate black hole horizon.  
To match onto the Lorentzian  solution all paths must have the    
``tapered  horn" shape, with $\tilde{N}^2 = 0$ at the degenerate horizon.
Since the horizon is an infinite proper distance from the rest of the space-time,  
there is no need to worry  
about conical singularities, and therefore no need to fix the pressure.    
Instead we leave $\sigma_{ij}$ fixed.    
    
Since the ultracold spacetimes are limits of the other cases,  
we argue that where there are no natural boundaries (and thus we 
considered them to be at infinity) the $\tilde{N}$ must still be fixed at 
these inserted boundaries.  
Again where the space-time is not closed there is no need to fix the 
pressure since there is no chance of  
a conical singularity at those points.    
    
To summarize, for the lukewarm, cold, and Nariai cases the    
imposition of regularity is equivalent to fixing the    
temperature, and therefore demanding that the created spacetimes be in    
thermal equilibrium. At the same time, we know that black holes are    
uniquely characterized by mass, angular momentum, and charge, so it is    
reasonable to demand of our interpolations that they have fixed angular    
momentum and charge. Fixing the mass is precluded since we must fix the    
temperature for the reasons discussed above.     

The only boundary term of (\ref{variation}) that we have now not
considered is the one on the $\Sigma_2$ surface that fixes $h_{ij}$. 
This is the natural quantity to fix in order to ensure that the paths will
match onto the Lorentzian solutions. Thus, there is no need to add a
boundary term in this situation. 
 
As noted earlier, creation rates for these spacetimes will be    
proportional to the action of their corresponding instantons. As it    
stands, those rates are calculated    
only up to a normalization factor (see the discussion in section    
\ref{formalism}). Rather than calculate this multiplicative    
factor, we shall calculate the probability of their creation relative     
to that of deSitter space. This is given by   
\be P = \exp{(2I_{deSitter}-2I)},    
\ee    
where $I$ is the action of the instanton, and $I_{deSitter}$ is the    
action of an instanton mediating the creation    
of deSitter space. Conventionally, this probability may also interpreted as    
the probability that deSitter space will tunnel into a given black hole 
space-time \cite{boussochamblin}.    
    
The space-like hypersurfaces of the spacetimes that we have considered are    
all of finite volume. In that case it is conventional    
\cite{mann,hawkingross}    
to interpret them as having constant energy (even though we have not    
explicitly fixed this quantity with boundary conditions). Then, the    
canonical partition functions that we have considered would be equivalent    
to the microcanonical partition function and as is standard in    
thermodynamics we may calculate entropies as,    
\be   
 \label{entropy}   
S = \ln{\Psi^2} = - 2I.    
\ee    
    
With these factors in mind we turn to an evaluation of the actions.    
    
\subsection{Evaluating the Actions}   
Based on the above considerations, we see that the basic action that keeps    
the angular momentum, magnetic charge, and boundary lapses fixed is   
\bea   
I_{magnetic} &=& I + \Delta I_{AngMom} \nn \\   
                  &=& \left( \mbox{ terms that vanish for stationary solutions }    
\right) \nn \\    
        &&  +\int_B d^3 x  \sqrt{\sigma} N (\varepsilon^{GR} +    
\varepsilon^{EM}), 
\eea   
where $I$ is the action (\ref{decomp}). 
In the above and in the following we have written the lapse in its    
Lorentzian form $N$ rather than its ``Euclidean'' form $\tilde{N}$. 
If we are evaluating one of these actions for an instanton, we    
substitute $\tilde{N}$ for $N$ in these expressions.   
   
We note that in all of the cases that we consider, $N \varepsilon^{GR} =    
0$ on the boundaries where we evaluate it. Further, on those boundaries   
$N \varepsilon^{EM}$ turns out to be proportional to $E_0^2 $. Thus, for    
magnetic instantons,     
\be   
I_{magnetic} = 0.   
\ee   
Of course extra boundary terms (the pressure terms) will have to be added 
to $I_{magnetic}$    
for most of our instantons, so the total action will not be zero. We will    
come to these terms in a moment.   
   
First however, we note that the basic action that keeps the angular    
momentum, electric charge, and boundary lapses fixed is   
\bea    
I_{electric} &=& I + \Delta I_{AngMom} + \Delta I_{electric} \nn \\    
                    &=& \left( \mbox{ terms that vanish for stationary solutions }    
                                       \right) \nn \\    
                     && + \int_B d^3 x N \sqrt{\sigma} N (\varepsilon^{GR} +    
\varepsilon^{EM}) - \frac{1}{\kappa} \int_M d^4 x \sqrt{-g} F^2.    
\eea    
Since we will only be evaluating actions for instantons (whose    
accompanying electromagnetic fields are solutions of the Maxwell    
equations), we have used Stoke's theorem to transform $\Delta
I_{electric}$ (\ref{deltaElectric}) into a bulk term which is easier to
evaluate. Now, as noted above $N \varepsilon^{GR} = 0$    
on the boundaries that we consider. Therefore, the basic action for electric    
instantons is   
\be   
I_{electric} = \int_B d^3 x  \sqrt{\sigma} N  \varepsilon^{EM}    
- \frac{1}{\kappa} \int_M d^4 x \sqrt{-g} F^2.    
\ee   
As we shall see, in all of the cases that we consider these two terms will    
evaluate to be    
equal in magnitude but opposite in sign and so cancel each other out,    
leaving us with $I_{electric} = 0$ again. Once again, the action will only    
be non-zero due to additional boundary terms.    
   
These additional terms arise because it also is necessary to fix the 
pressure    
$p$ whenever there is a non-degenerate single horizon. 
In conjunction with fixing the lapse $N$ this ensures regularity of all paths 
at the horizon, sufficient to fix the temperature there as noted above. 
 
Hence wherever there is a single, non-degenerate horizon, we must add the 
boundary term
\be\label{delipres} 
\Delta I_{pressure}  = - \frac{1}{2} \int_{B_H} d^3 x \sqrt{\sigma} N p    
\ee   
to the action,  where $B_H$ is the appropriate boundary corresponding to 
the horizon crossed with the time coordinate over the range  
$[0, \frac{P_0}{2}]$.  From (\ref{pressure}) we have 
$ N p \rightarrow  \frac{2}{\kappa} n^k \partial_k N$, and by    
(\ref{periodfix}), $n^k \partial_k N = \frac{2 \pi}{P_0}$   
on a non-degenerate horizon. Evaluating (\ref{delipres}) by 
integrating over the instanton yields 
 
\be   
\Delta I_{pressure} = - \frac{\cA_H}{8} 
\ee   
where $\cA_H$ is the surface area of the surface $\Omega_{H;2}$.    
Hence  the action is equal to $-\frac{1}{8}$ times the sum of the areas of  
the non-degenerate    
horizons for all classes of instantons. 
 
We give specific values for these quantities below (as well as the    
values of the cancelling terms in the electric actions).   
   
\underline{Lukewarm Action:}     
In this case, there are non-degenerate cosmological and outer black hole    
horizons. Therefore the total action of the magnetic instantons is   
\be   
I_{MLW} = - \frac{\cA_c + \cA_h}{8} = - \frac{\pi (r_c^2 +  
a^2)}{2\chi^2}    
- \frac{\pi(r_h^2 + a^2)}{2 \chi^2},   
\ee   
where $\cA_c$ and $\cA_h$ are respectively the areas of the cosmological    
and outer black hole horizons in the Lorentzian solution.   
   
For the electric lukewarm instantons, we note that   
\be   
\int_0^{\frac{P_0}{2}} d\tilde{t}  \int_{\Omega_{\tilde{t}}} d^2 x    
\sqrt{\sigma} \tilde{N} \varepsilon^{EM} = \frac{1}{\kappa} \int_M d^4    
x \sqrt{\tilde{g}} \tilde{F}^2  = \frac{P_0 E_0^2}{2 \chi^2} \left(    
\frac{r_c}{r_c^2 + a^2} - \frac{r_h}{r_h^2 + a^2} \right),   
\ee   
and so the $I_{electric} = 0$ as asserted, yielding  
\be   
I_{ELW} = - \frac{\pi (r_c^2 + r_h^2 +  2a^2)}{2\chi^2}    
\ee   
for the total electric lukewarm action  as well.   
   
\underline{Nariai Actions:}   
Again  there are two non-degenerate horizons, this time at $\rho = \pm 1$.    
Therefore the total action of the magnetic Nariai instantons is   
\be   
I_{MN} = - \frac{\cA_{\rho=-1} + \cA_{\rho=1}}{8} =  
- \frac{ \pi (e^2+a^2)}{\chi^2},   
\ee   
where $\cA_{\rho = \pm 1}$ is the area of the horizon at $\rho = \pm 1$.    
We note that for the Nariai $\cA_{\rho=1} = \cA_{\rho=-1}$.   
   
For the electric Nariai instantons,   
\be   
\int_0^{\frac{P_0}{2}} d\tilde{\tau} \int_{\Omega_{\tilde{\tau}}} d^2 x    
\sqrt{\sigma} \tilde{N} \varepsilon^{EM} = \frac{1}{\kappa} \int_M d^4    
x \sqrt{\tilde{g}} \tilde{F}^2  = - \frac{P_0 E_0^2 (e^2-a^2)}{\chi^2    
(e^2+a^2)}   
\ee   
and so the $I_{electric} = 0$ as claimed. The total electric Nariai    
action is 
\be   
I_{EN} = - \frac{ \pi (e^2+a^2)}{\chi^2} ,  
\ee  
equivalent to the magnetic case.

\underline{Cold Actions:}   
Here there is only one non-degenerate horizon, and so 
\be   
I_{MC} = - \frac{\cA_c}{8} = - \frac{\pi (r_c^2 + a^2)}{2 \chi^2},   
\ee   
where $\cA_c$ is again the area of the cosmological horizon. Note that  
we consider the regions of the instanton to lie between $r = r_c$  
and the degenerate horizon at $r = r_h$.    
   
For the electric cold instantons the two terms of $I_{electric}$ take the    
same values that they did in the lukewarm case, and so $I_{electric}= 0$ as 
promised.  Hence the total electric cold action is given by 
\be   
I_{EC} =  - \frac{\pi (r_c^2 + a^2)}{2 \chi^2} 
\ee   
as well.   
   
\underline{Ultracold I Actions:}   
Again there is only a single nondegenerate horizon, this time at $R=0$.    
The action of the magnetic ultracold I instanton is  then 
\be   
I_{MUCI} = - \frac{ \cA_{R=0}}{8} = -\frac{\pi (e^2+a^2)}{2 \chi^2}.   
\ee   
where this time the region we consider in our calculation is between $R=0$  
and    
$R = R_+$. We shall take the limit $R_+ \rightarrow    
\infty$ at the end of our calculation  
in order to include the whole instanton.   
   
For the electric ultracold I instantons,    
\be   
\int_0^{\frac{P_0}{2}} d\tilde{T} \int_{\Omega_{\tilde{T}}} d^2 x    
\sqrt{\sigma} \tilde{N} \varepsilon^{EM} = \frac{1}{\kappa} \int_M d^4    
x \sqrt{\tilde{g}} \tilde{F}^2  = - \frac{P_0 E_0^2 (e^2-a^2)}{2    
\chi^2 (e^2+a^2)} R_+   
\ee   
so $I_{electric} = 0$ as asserted, and   
\be   
I_{EUCI} = -\frac{\pi (e^2+a^2)}{2 \chi^2} 
\ee   
as well.   
   
\underline{Ultracold II Actions:}   
There are no horizons whatsoever for this case, and so 
\be   
I_{MUCII} = 0,   
\ee   
irrespective of the chosen period $P_0$ of the ``time'' coordinate.   
We consider the region between  $R=R_-$ and $R=R_+$, taking the    
limits $R_- \rightarrow - \infty$ and   $R_+  \rightarrow \infty$  
at the end of the calculation to include the whole instanton.   
   
For the electric ultracold II instantons,   
\be   
\int_0^{\frac{P_0}{2}} d\tilde{T} \int_{\Omega_{\tilde{T}}} d^2 x    
\sqrt{\sigma} \tilde{N} \varepsilon^{EM} = \frac{1}{\kappa} \int_M d^4    
x \sqrt{\tilde{g}} \tilde{F}^2  =  -\frac{P_0 E_0^2 (e^2-a^2)}{2    
\chi^2 (e^2+a^2)} (R_+-R_-),    
\ee   
yielding $I_{electric} = 0$ and so 
\be   
I_{EUCII} = 0,   
\ee   
as well.   
   
In figure \ref{creationrates} we plot the above actions as a fraction of    
the action of the instanton creating deSitter space with the same    
cosmological constant. For all cases $I, I_{dS} < 0$ and from the diagram    
we see that $|I| < |I_{dS}|$. Then $I_{dS} - I < 0$ and we see that each of    
the space-times considered above is less likely to be created than deSitter    
space. Note that the Nariai space-time is the most likely to be created  
provided the parameter values are such that the instanton exists, 
while the cold space-time is the least likely to be created.    
As we might expect on physical grounds, smaller and more slowly rotating    
holes are more likely to be created than larger and more quickly rotating    
ones. As $\frac{a}{M} \rightarrow 0$ and $M \rightarrow 0$,    
the creation rates approach those of deSitter space.   
   
Assuming that the space-times are at least quasi-static    
(see section \ref{quasi-static} below), using equation    
(\ref{entropy}) we see that the entropy of these space-times is equal to one-   
quarter of the sum of the areas of non-degenerate horizons bounding the    
Lorentzian region of the space-time.   
Consistent with \cite{hhr} and \cite{mann}, the degenerate horizon in the    
cold case does not contribute to the entropy of the cold space-time.    
   
\begin{figure}    
\centerline{\psfig{figure=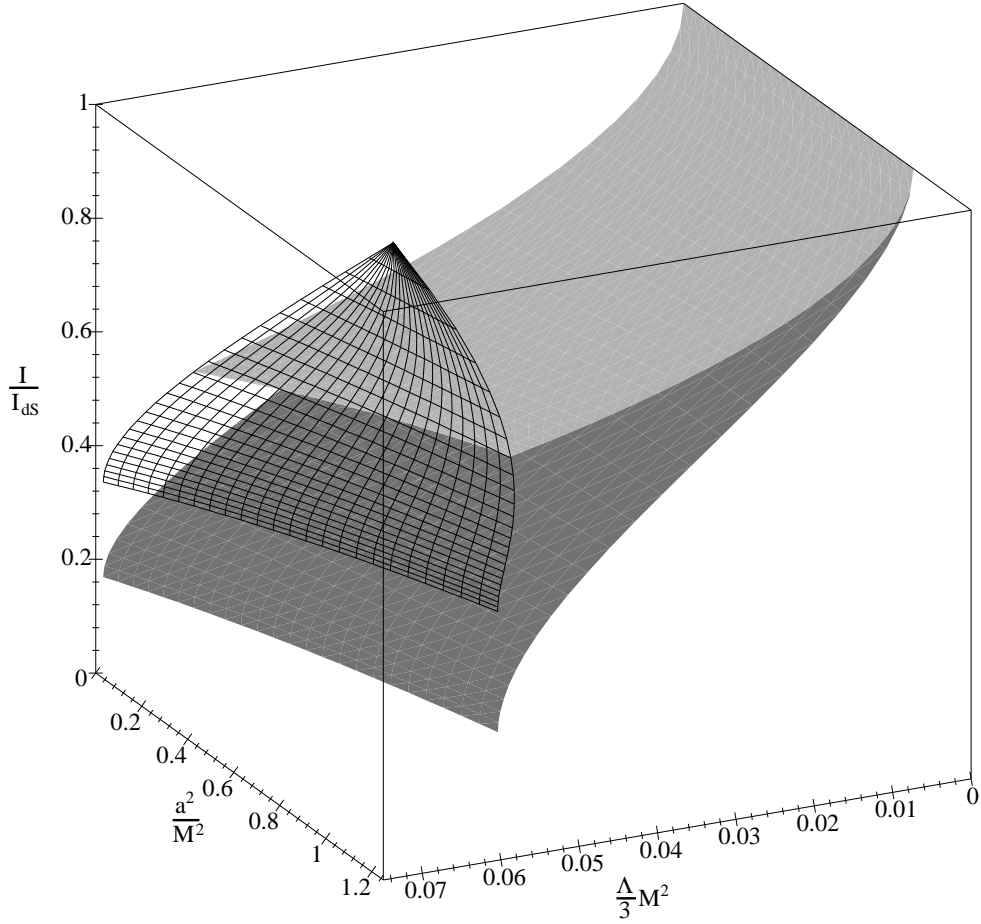,height=13cm}}
\caption{The actions for the charged and rotating lukewarm, cold, and
Nariai instantons. The instantons are parameterized by $\frac{a^2}{M^2}$
and $\frac{\Lambda}{3}M^2$. Their actions are  plotted as a fraction of
the action $I_{dS}$ which is the action of the instanton creating pure 
deSitter space with the same
cosmological constant. The Nariai instantons are the meshed sheet, the
lukewarm instantons are the lighter grey sheet, and the cold instantons
are the darker grey sheet.The ultracold I instanton actions may be found
at the
``bottom'' end of the cold sheet, while the ultracold II instanton actions
are zero.}    
\label{creationrates}    
\end{figure}    
    
\subsection{Comparison to extant calculations}    
    
The approach we have taken in computing the instanton actions  
differs from those carried out for non-rotating black holes \cite{mann}. We 
pause to comment on the relationship between these cases.  
 
In ref. \cite{mann} the fact that the instantons are closed and smooth at the  
points    
corresponding to the non-degenerate horizons was taken to mean that no    
boundary terms need be considered there, implying that  
the basic action used for the lukewarm and Nariai    
instanton should be 
\bea    
I_{old} = - \frac{1}{2 \kappa} \int_M d^4 x \sqrt{-g} \left( \cR -    
2\Lambda - F^2\right)    
- \frac{1}{\kappa} \int_{\Sigma} d^3x \sqrt{h} K,    
\label{oldact}    
\eea    
which is our action (\ref{basicAction}) with the boundary term    
\bea    
\frac{1}{\kappa} \int_{B} d^3x \sqrt{-\gamma} \Theta   
\eea    
added on.  It is easy to see that this term is equivalent to the 
pressure term   
\bea    
-\int_{B} d^3x N \sqrt{\sigma} \frac{p}{2} = -\int_{B} d^3x N    
\sqrt{\sigma}    
\left[ \frac{n^i \partial_i N}{N} - \frac{k}{2} \right],    
\eea    
evaluated on the equivalent horizons. Noting that    
$\Theta = k - \frac{n^i \partial_i N}{N}$, and    
$k = - \frac{1}{2 \sqrt{h_{rr}}} \partial_r \ln\sigma$,     
and $\frac{1}{\sqrt{h_{rr}}} \rightarrow 0$ at each horizon, we see that    
on the horizons $\Theta = -\frac{p}{2}$, and so in the non-rotating case
our approach is equivalent to that of \cite{mann}.   

For the cold case $k$ still vanishes on the boundary and so the inclusion of    
the $\Theta$ term in \cite{mann} is the equivalent of the omission of the    
pressure term in our calculations. Finally, in the ultracold cases $k=0$    
everywhere and so once more the omissions/inclusions are equivalent.   
   
For electric instantons in both calculations electromagnetic boundary terms    
are added to the action to fix the electric charge for all paths considered in    
the path integral. Further, in both calculations for solutions to the Maxwell    
equations, these boundary terms may be converted into the    
$F^2$ bulk term that we have used.    
    
Although for non-rotating instantons our approach    
is equivalent to earlier ones for evaluating the actions,    
differences arise when we include rotation. In earlier approaches 
\cite{empair,hhr,mann,othercosmo,strpair,dompair} there was no provision
made for fixing 
the angular  
momentum. 
 
The action differs by the term $\Delta_{AngMom}$ and its omission 
is tantamount to working with an incorrect    
thermodynamic ensemble. Evaluating the action of rotating  
instantons with (\ref{oldact}) will not yield the preceding relationships 
linking surface areas, actions, and entropies.  
Indeed, using (\ref{oldact}) the creation rate of    
rotating black holes is enhanced relative to that of    
non-rotating black holes and with an appropriate choice of physical    
parameters may be made arbitrarily large.  
    
Recently Wu has considered the creation of a single black hole through the  
use of    
(slightly different) KNdS instantons \cite{wu}. Although we concur with 
the modifications that must be made to the action in order to properly  
take angular momentum into account, we disagree with the description of 
black hole creation presented. Creation of a single    
hole, as Wu considers, will not conserve angular momentum and  
electric/magnetic  
charge. Furthermore the instantons considered in that paper do not properly  
match to real Lorentzian solutions for two reasons. In the first place there  
are no periodic identifications of the universal covering space of the  
basic KNdS solution that can be made such that $t=\mbox{constant}$  
hypersurfaces will contain only a single black hole. The smallest number  
of black holes that may be  contained are     
the two that we have discussed in this paper. Second, as we have argued     
earlier, an analytic continuation of $a$ to $ia$ and $E_0$ to $iE_0$ will    
mean in  general that an instanton generated from a classical solution  
will not properly match onto that classical     
solution: there will in general not even be the correct number of    
horizons available in the instanton to match onto the Lorentzian solution, 
and extrinsic curvatures, induced metrics, and induced electromagnetic    
fields will not match across a $t=\mbox{constant}$ surface.      
    
\subsection{Issues of Equilibrium II}    
\label{quasi-static}    
   
Finally we return briefly to the discussion in section \ref{equil}.  
Recall that the spacetimes of the pair-created black holes are in    
thermal equilibrium, but not in equilibrium with respect to the    
charged particle creation and super-radiance effects. These will    
cause the black hole spacetimes to discharge and/or spin down.  
 
This problem is also    
present (at least in principle) in previous work on charged black hole    
pair-creation \cite{empair,hhr, mann,dompair,bousso}) since in those
cases     
charged black holes tend to discharge via charged particle creation     
effects.        
    
The first, and more conservative response to this situation, 
is to argue that even if the    
created spacetimes are not static (in the thermodynamic sense), then    
perhaps their evolution is slow enough that they may be viewed as    
quasistatic. In section \ref{equil} we saw that there is some evidence for    
this point of view in the literature, though admittedly our current    
situation has not been explicitly addressed. It seems clear however that at    
least some finite class of the spacetimes that we have considered will be    
close enough to equilibrium that the calculations will be correct to the    
first order of approximation. In a future paper we will explore this  
quantitatively. 
    
An alternate (and more radical) response is to argue that     
only thermal equilibrium    
(in the sense of equal temperatures at the horizons) is needed for pair    
creation. Here we would argue that while the requirement of thermal    
equilibrium arises naturally from the smoothness conditions on the    
instantons, the extra requirements of full thermodynamic equilibrium do    
not arise naturally from such conditions and are instead imposed from  
outside. Of course on physical grounds we would expect that full    
thermodynamic equilibrium would be required, but it at least seems    
possible that this might not be the case. For now we will leave this issue    
open.

\section{Discussion}   
   
We have demonstrated the physical process of creation of static black   
hole pairs via cosmological vacuum energy may be extended to the  
creation of stationary black hole pairs. Although the calculation is 
somewhat more   
subtle and complicated, the basic   
results continue to hold qualitatively. Just as there are static   
lukewarm, cold, Nariai, and ultracold instantons describing pair-creation   
in a cosmological (or for that matter electromagnetic \cite{empair})  
background,   
so also are there the same classes   
of instantons for rotating black hole pairs. Furthermore, the entropy of such   
spacetimes continues to be proportional to the sum of the areas of the   
horizons in the standard manner, and pair creation rates continue to be   
proportional to the exponential of those entropies and suppressed relative   
to the creation of a pure deSitter space.   
   
In order to describe this process we have had to depart from the usage 
of purely real Euclidean instantons and consider complex instantons. 
This followed from a consideration of the standard matching   
conditions required in the Euclidean gravity formalism which specify that   
instantons must smoothly match to their corresponding Lorentzian  
solutions. Demanding that rotating instantons be real implies 
they will no longer match onto their Lorentzian counterparts.  
Although it is somewhat unusual to   
introduce complex metrics, our results are consistent with the interpretation 
of the functional integral formalism for black hole thermodynamics 
discussed  
in ref.\cite{BMY}.

A second departure from the standard techniques arises due to  the 
boundary terms we must add  when rotation is present. In pair-creation 
calculations 
for non-rotating black holes, one uses the basic   
Einstein-Maxwell action (\ref{basicAction}), and we have seen that this   
action is equivalent to the one that we have used. However it must be 
modified by additional boundary terms when rotation is present in order 
to appropriately fix angular momentum on the matching surface, somewhat 
analogous to the situation in which boundary terms must be added in the 
electric case in order to maintain electromagnetic duality 
\cite{hawkingross,brownEM}. 
Usage of (\ref{basicAction}) for rotating cases yields the unphysical   
result that pair-creation of rotating holes is enhanced rather   
than suppressed relative to deSitter space.  
   
Finally, we have seen that thermal equilibrium does not   
necessarily correspond to thermodynamic equilibrium. The question as to   
which is the correct requirement to put on spaces that are to be created   
by quantum tunnelling   
has been left open. Geometrically it would appear that only   
thermal equilibrium is required, but on physical grounds we would expect   
full thermodynamic equilibrium to be required. We have suggested that  
even if full thermodynamic equilibrium is required, then at least a class of   
the spacetimes that we have discussed will be close enough to equilibrium   
to be considered quasistatic, in which case the approximate entropies and   
pair creation rates should still be correct.

\section*{Acknowledgements}   
   
This work was supported by the Natural Sciences and Engineering  
Research Council  
of Canada.

\newcommand{\PR}{Phys. Rev.}     
\newcommand{\PRL}{Phys. Rev. Lett.}     
\newcommand{\CQG}{Class. Quantum Grav.}     
\newcommand{\CMP}{Commun. Math. Phys.}     
\newcommand{\NP}{Nucl. Phys.}     
\newcommand{\PL}{Phys. Lett.}


\end{document}